\documentclass[useAMS,usenatbib]{mn2e}

\usepackage{graphicx}
\usepackage{lastpage}

\title[]
  {Dark matter equation of state from rotational curves of galaxies}
\author[J. Barranco et al.]
  {J.~Barranco,$^1$\thanks{jbarranc@fisica.ugto.mx}
  A.~Bernal,$^2$\thanks{argelia\_bernal@uaeh.edu.mx} D.~N\'u\~nez,$^3$\thanks{nunez@nuclares.unam.mx. Affiliated to IAC.}\\
 $^1$Departamento de F\'isica, Divisi\'on de Ciencias e Ingenier\'ias, Campus Le\'on,\\ 
Universidad de Guanajuato, Le\'on 37150, M\'exico.\\ 
$^2$\'Area Acad\'emica de Matem\'aticas y F\'isica, Universidad
Aut\'onoma del Estado de Hidalgo, \\Carretera Pachuca-Tulancingo Km. 4.5, C.P. 42184,
Pachuca, Hidalgo, M\'exico.\\
$^3$Instituto de Ciencias Nucleares, Universidad Nacional
 Aut\'onoma de M\'exico, \\Circuito Exterior C.U., A.P. 70-543,
M\'exico D.F. 04510, M\'exico.}

\pagerange{\pageref{firstpage}--\pageref{lastpage}} \pubyear{2014}

\begin{document}

\label{firstpage}

\maketitle

\begin{abstract}
In this work we model galactic halos  describing the dark matter
as a non zero pressure fluid and derive, not impose, a dark matter equation of state by using observational 
data of the rotation curves of galaxies. In order to 
reach hydrostatic equilibrium, as expected for the halo, it is mandatory that dark fluid's pressure should not be zero. 
The equation of state is obtained by solving the matter-geometry system of equations assuming different dark 
matter density or rotational velocity profiles. 
The resulting equations of state are, in general, different to a barotropic equation of state. The free parameters  
of the equation of state are fixed by fitting the observed rotational velocities of a set of galaxies. 
\end{abstract}


\section{Introduction}

Astrophysical observations of the red shift of stars in galaxies, 
of the cosmic microwave background anisotropies, 
of the deflection of the light of distant galaxies, among others, 
are all explained by the presence of dark matter, which has not shown any interaction with 
the baryonic matter except the gravitational one. See Bertone (2005), Dodelson (2003) and Binney \&
Tremaine (2010), for a thorough discussion on dark matter.

On the other hand, direct detection of dark matter demands it to 
have some interaction with the rest of particles of the standard model,
see Smith (1990) and Gaitskell (2004).
These two research lines seem to have contradictory points of view on the properties of the
dark matter. In order to attempt to conciliate these seemingly opposite strategies, one can research on
the possibility that the astrophysical description of dark matter could relax the pressureless hypothesis, 
and allow some interaction with itself and/or with the baryonic matter. 

For instance, at the cosmological level, M\"uller (2005)
proposed a barotropic relation for the dark fluid, $p=\omega_{\rm DM}\,\rho$,
instead of considering the usual dust case ($\omega_{\rm DM}=0$). The
author shows that the combined
analysis of data from anisotropies of the cosmic microwave background, supernovae Ia and matter power spectra
constrain $\omega_{\rm DM}$ to lay within the interval $[-1.5,1.13]\,\times 10^{-6}$
(see also Calabrese et al. (2009), and also Avelino et al. (2012)), implying that a small, but non-zero
dark matter pressure within the $\Lambda$CDM model, remains to be consistent with the observations.
 Furthermore, in order to avoid that intermediate mass black holes grow to values
beyond the observed upper limits due to dark matter accretion, it seems mandatory that dark matter, modeled as
a fluid,  should have a non 
zero pressure, as pointed out in Pepe et al (2011). The same conclusion applies for
Super massive black holes Guzman \& Lora-Clavijo (2011).
Note that the dark pressure is not likely to have an electromagnetic nature, so that it should 
have a different origin, raising 
as a kinetic pressure, for instance, or be of a weak interaction type.

Since one of the strongest evidence for dark matter comes from the rotational curves, 
it is natural to relax the pressureless hypothesis at galactic scales when describing the dark matter as a fluid.
In the present work, we show that if dark matter is described by a perfect isotropic 
fluid, assuming a spherical symmetric and 
static space-time, 
a functional relation between pressure  and 
energy density emerges using only the rotational velocity profiles. 
If the isotropic assumption is relaxed, 
then it is needed another observation to determine the equation of state. Such observation can be 
the gravitational lensing, see for instance 
Faber \& Visser (2006), N\'u\~nez et al (2010), and Serra \& Romero (2011) for a further discussion on
the topic.

There are other works where both luminous and non-luminous matter is considered in axial symmetric space times with no pressure fluids, that is, models where the halo is rotating, see for instance (Cooperstock \& Tieu 2005a; Cooperstock \& Tieu 2005b; Cooperstock \& Tieu 2006; Cooperstock \& Tieu 2007; Cooperstock \& Tieu 2008; Carrick \& Cooperstock 2012)). It maigh be interesting to study in a future work those models and test the derivation of the equation of state for several halo models, as well as the Newtonian approximation in that case.

In order to show how to obtain the equation of state for the dark matter, we have
organized the article as follows: In Section \ref{perfect_fluid} we present the general relativistic equations 
for an isotropic perfect fluid in a spherically symmetric and static 
spacetime, and discuss how one of the metric coefficients is completely 
determined by the rotational curve profile. Hence, the resulting system of three equations involves three unknowns: the
dark matter  fluid pressure and density, and the other metric coefficient, that is, the system is closed.

We then consider the Newtonian limit of those equations, since
this problem is more transparently analyzed in that limit. In the Appendix \ref{apendice} we show that
the equation of state obtained via the Newtonian 
limit is an excellent approximation to the one obtained in the general relativistic treatment. Therefore, we use the
Newtonian formalism in the rest of the paper, for clarity purpose.
In section \ref{ajustes} we use the rotational velocity profile proposed in Persic \& Salucci \& Stel (1996),
PSS profile,
to obtain the equation of state 
and we fixed the free parameters by fitting with them the observed velocities of a set of galaxies.
In section \ref{density_profiles} we derive the equation of state, starting from the density profile instead.
We have used the Pseudo-isothermal density profile (subsection \ref{pseudoisoterma}), the Einasto density profile, Einasto (1965) 
(subsection \ref{Einasto}),
the Navarro-Frenk-White, NFW, density profile, Navarro \& Frenk \& White (1997) (subsection \ref{NFW_profile}) 
and the Burkert density profile, Burkert (1995) (subsection \ref{Burkert}).
In section \ref{discussion} we make a comparison between all the equations of state we have derived for each
model and discuss some implications for dark matter detection and possible follow up research.

\section {Perfect fluid halo} \label{perfect_fluid} 
Although the halo could have been directly considered as a Newtonian system, we used the best available theoretical tool, i.e. general relativity, given the unknown nature of the dark matter, and then test the accuracy of the Newtonian description.
We consider a static and spherically symmetric space-time in General Relativity, 
described by the line element:
\begin{equation}
ds^2=-e^{2\Phi/c^2} \,c^2\,dt^2+ \frac{dr^2}{1-\frac{2\,G\,m}{c^2\,r}}+r^2(d\theta^2+\sin^2\theta\,d\varphi^2)\,,
\label{eq:lel}
\end{equation}
where the gravitational potential $\Phi(r)$ and the mass function $m(r)$ are functions of the
radial coordinate only.

We will take advantage of the fact 
that one of the geometric potentials can be determined by the observations. Indeed, it 
can be obtained an expression relating the tangential velocities of test particles, $v_t$,
following stable circular orbits in this spacetime, with the gravitational potential $\Phi$, 
by means of solving the geodesic equations for test particles in circular stable orbits 
(see Matos et al (2000), Matos et al. (2000a), Cabral (2002) and Rahaman et al. (2010) for details on the derivation).
Such expression is
\begin{equation}
\frac{\Phi'}{c^2}=\frac{\beta^2}{r}\,, \label{eq:phi_v}
\end{equation}
where $\beta^2=\frac{v_t^2}{c^2}$. 

The same expression, Eq. (\ref{eq:phi_v}),
is obtained in the Newtonian description 
by simply equating the  gravitational force with the centrifugal one for particles in circular orbit.

Thus, given a rotational velocity profile $\beta(r)$, 
the system of a self-gravitating isotropic fluid in a space time with line element Eq. (\ref{eq:lel}), 
is left with three unknowns: one geometric, the mass function, and two related to the dark fluid, the pressure and the 
energy density. The system of equations are two Einstein's equations and the conservation equation,
that is, the system is closed. 
Indeed, using the relation (\ref{eq:phi_v}), the resulting equations
are:
\begin{eqnarray}
n'&=&3\,x^2\,\bar{\rho}\,, \label{eq:nRG}\\
\left(1-q\,\frac{2\,n}{x}\right)\,\frac{\beta^2}{x} - q\,\frac{n}{x^2}&=&3\,q\,x\,\bar{p}, \label{eq:pRG}\\
\bar{p}' + \left(\bar{p}+\bar{\rho}\right)\,\frac{\beta^2}{x}&=&0, \label{eq:edoRG}
\end{eqnarray}
where prime stands for derivative with respect to $x$ and 
we have used convenient dimensionless variables $n$ and $x$ for the mass function and the 
radial distance respectively, such that $m=M_\star\,n$ and $r=R_\star\,x$, where $M_\star$ 
and $R_\star$ are the characteristic scales for mass and distance of the
system under study. 
We also define a characteristic density, $\rho_\star=M_\star/(4/3)\pi\,{R_\star}^3$,
and a characteristic pressure: $p_\star=\rho_\star\,c^2$, so that $\rho=\rho_\star\,\bar{\rho}$, and $p=p_\star\,\bar{p}$, 
with $\bar \rho, \bar p$ dimensionless functions.
In the Eq. \ref{eq:pRG} it was also introduced the dimensionless parameter 
$q=M_\star\,G/c^2\,R_\star$, which meassures the characteristic compactness of the system.

Let us consider the Newtonian limit of 
Eqs.~(\ref{eq:nRG}-\ref{eq:edoRG}). 
Notice that Eq.~(\ref{eq:nRG}) does not change; in the Newtonian limit it can be considered
as the definition of the mass function in terms of the density (see Landau and Lifshitz (1971), 
and Misner and Sharp (1964)). The second Einstein's equation, in the 
weak field limit and considering that the pressure is negligible, gives a direct relation between the
gravitational potential and the mass function, $\frac{\Phi'}{c^2}=q\frac{n}{x^2}$, hence,  in the
Newtonian description the two geometric functions are related, in contrast to the  general relativity case, where 
the geometric functions are completely independent functions. 
Finally, in the conservation equation, the Newtonian limit implies
that the pressure is much less than the density, so neglecting the pressure as compared to the density in the second term of 
Eq. (\ref{eq:edoRG}), we obtain the usual Euler
equation in equilibrium. Thus, the Newtonian limit of the relativistic equations Eqs.~(\ref{eq:nRG}-\ref{eq:edoRG}) are:
\begin{eqnarray}
n'&=&3\,x^2\,\bar{\rho} \,,\label{eq:Nrho}\\
\beta^2\,x&=&q\,n, \label{eq:Nn}\\
\bar{p}' + \bar{\rho}\,\frac{\beta^2}{x}&=&0\,. \label{eq:Nt}
\end{eqnarray}

In this way, given the rotational velocity profile, $\beta^2(x)$, the gravitational potential, $\Phi$,
and the mass, $n$, are determined directly. Next, the density is determined straightforward, and one
has only to solve  one differential equation for the pressure, namely:
\begin{equation}
\bar{p}' + \frac{\beta^2}{3\,q\,x^3}\,\left(\beta^2\,x\right)'=0\,. \label{eq:pn_only}
\end{equation}

Once the pressure and the density are known, it is possible to determine the equation of state. 
Notice that $\bar p=0$ is not a solution of Eq. (\ref{eq:pn_only})  for a given non zero rotational 
velocity profile. It is mandatory that the dark fluid should have
pressure. 

In the case when the density profile is the starting point
instead of the rotational velocity profile, 
it is also possible to determine all the functions of the system and
again derive an equation of state. We will show examples of both cases in the next section.

The relativistic case is more involved, but an analogous procedure can be followed, and also obtained at
the end a differential equation for the pressure, such procedure is detailed in the appendix.
We stress the fact that the exact treatment is the relativistic one and we have used it 
to derive the corresponding equation of state, 
but we found that the relativistic treatment gives only negligible corrections to the Newtonian results. 
Thus, from now on we will work only in the Newtonian limit.

Even though the halo could have been directly considered as a Newtonian system, we used 
the best available theoretical tool, given the unknown nature of the dark matter, and then test the approximated
description.

\section {Dark Matter equation of state from rotational curves}\label{ajustes} 

\begin{center}
\begin{figure}
\includegraphics[width=0.5\textwidth]{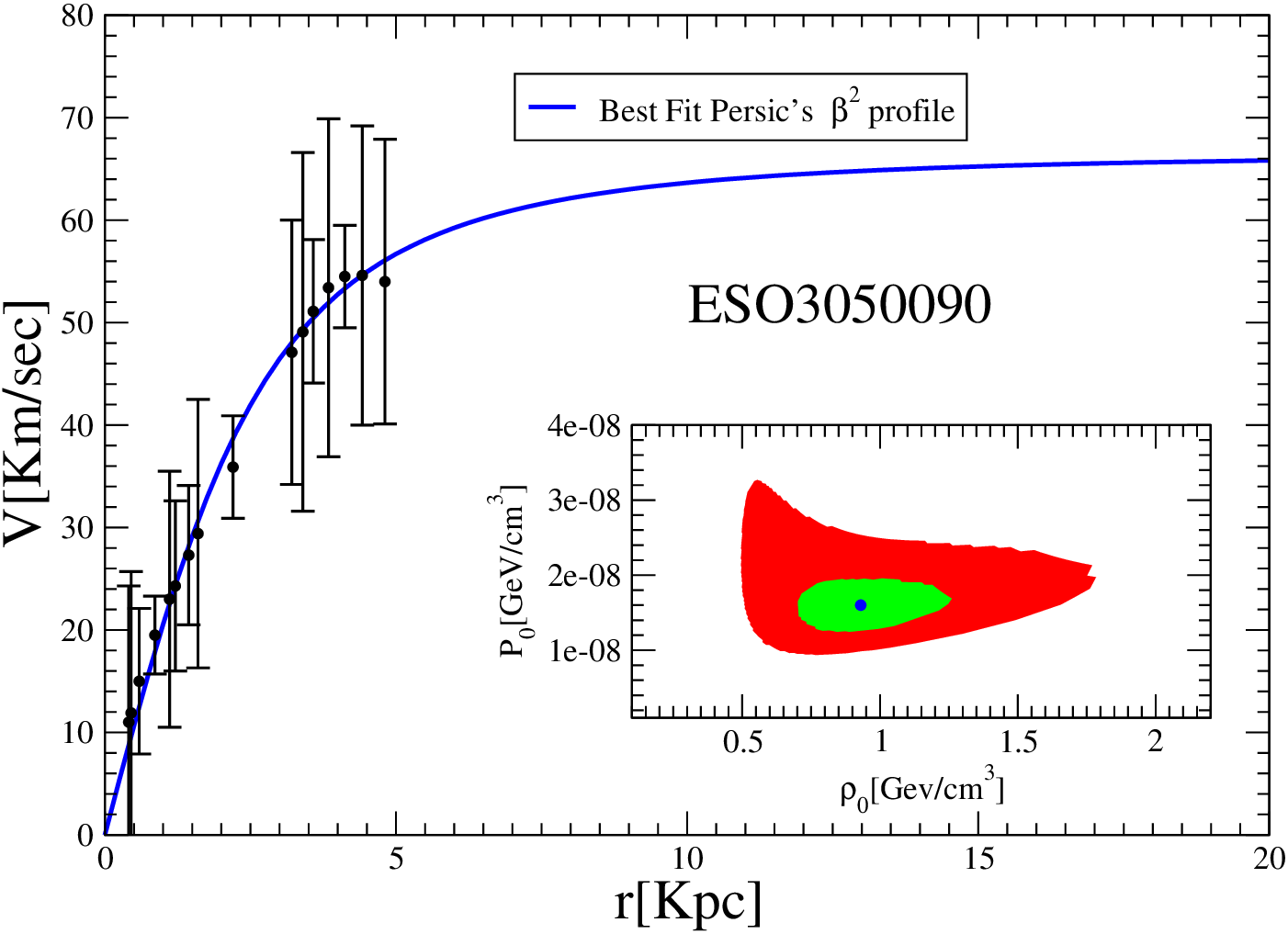}
\caption{Fit on the rotation curves for the galaxy ESO3050090. Red line corresponds to best fit point as 
expressed in table \ref{table1}. Inner figure:  Green region corresponds to 
$68$ confidence level and red region to $90\%$ confidence level values for $(\rho_0,p_0)$}\label{Fig1}
\end{figure}
\end{center}

There are several profiles which have been proposed to adjust the rotational velocities profiles
observed in the galaxies (see, e.g., Courteau 1997).

We will consider the velocity profile proposed by Persic \& Salucci \& Stel (1996) obtained by
adjusting 1023 galaxies which is given by:
\begin{equation}
\beta^2={\beta_0}^2\,\frac{x^2}{x^2+a^2}, \label{eq:beta_sal}
\end{equation}
where $\beta_0, a$ are constants. The first one, $\beta_0$, stands for the ratio of the terminal 
velocity to the speed of light, and $a$ determines how fast the velocity reaches a terminal value. 
Within the notation of Persic \& Salucci \& Stel (1996), 
$\beta_0$ is a function of the luminosity of the galaxy.
In our present work we will assume that both $\beta_0^2$ and $a$ are simply constants that can be fitted 
to observational data and from now we will refer to Eq. (\ref{eq:beta_sal}) as the PSS rotational velocity profile. 

Given $\beta^2(x)$ by Eq.~(\ref{eq:beta_sal}), it is straightforward to obtain 
from Eqs.~(\ref{eq:Nrho}-\ref{eq:Nt}) analytical expressions for the mass 
function, the density and the pressure:
\begin{eqnarray}
n&=&\frac{{\beta_0}^2}{q}\,\frac{x^3}{\left(x^2 + a^2\right)} \label{eq:n_sal_n} \\
\bar{\rho}&=&\frac{{\beta_0}^2}{3\,q}\,\frac{\left(x^2 + 3\,a^2\right)}{\left(x^2 + a^2\right)^2}, \label{eq:rho_sal_n} \\
\bar{p}&=&\frac{{\beta_0}^4}{6\,q}\,\frac{\left(x^2 + 2\,a^2\right)}{\left(x^2 + a^2\right)^2}\,, \label{eq:p_sal_n}
\end{eqnarray}
and it is then possible to obtain an analytical expression for the equation 
of state for dark matter:
\begin{equation}
\bar{p}=\frac{{\beta_0}^4}{48\,a^2\,q}\,
\left(
\frac{12\,a^2\,q}{{\beta_0}^2}\,\bar{\rho} - 
\left(1 - \sqrt{1 + \frac{24\,a^2\,q}{{\beta_0}^2}\,\bar{\rho}} \right)
\right).
\label{eq:equation of state1}
\end{equation}

Furthermore, the density and the pressure profiles as expressed in Eqs.~(\ref{eq:rho_sal_n},\ref{eq:p_sal_n})
allow us to find a relation between $\beta^2_0$ and $a$ as functions of the central pressure and the 
central density $\bar \rho_0$ and $\bar p_0$. Namely
\begin{equation}
a^2=\frac{3\,\bar{p}_0}{q\,{\bar{\rho}_0}^2}\,, \quad \beta_0^2
=\frac{3\,\bar{p}_0}{\bar{\rho}_0}\,.\label{eq:rho0p01}
\end{equation}
\begin{table}
\begin{tabular}{| c| c| c| c| c| c| }
\hline \hline
$Galaxy$ & $\beta_0\,10^{-4}$  & $a$ & $\rho(0)~ \frac{\mbox{GeV}}{\mbox{cm}^3}$ & $p(0)~\frac{\mbox{GeV}}{\mbox{cm}^3}$ & 
$\frac{\chi^2_{min}}{\mbox{d.o.f.}}$ \\
\hline \hline
F563 1       & 3.74  & 3.18  & 2.50    		 & $1.2 \times 10^{-7}$     & 0.13 \\
F568 3       & 3.71  & 4.53  & 1.22     		& $5.6\times 10^{-8}$   & 0.34 \\
F570 v1      & 4.78  & 2.55  & 6.40    		& $4.9\times 10^{-7}$  & 0.35  \\
F571 8       & 4.8  & 3.65  & 3.13       		& $2.4\times 10^{-7}$  &2.2 \\
F579 v1      & 3.74  & 1.29  & 15.30  		& $7.1\times 10^{-7}$  &0.08\\
F583 1       & 3.04  & 1.48  & 1.04    		&  $3.\times 10^{-8}$ & 0.01\\
F583 4       & 2.22  & 1.77  & 2.85    		&  $4.7\times 10^{-8}$ & 0.57\\
ESO140040    & 9.2  & 5.08  & 5.91		&$1.7\times 10^{-6}$  & 0.45\\
ESO2060140   & 4.00  & 2.16  & 6.22 	& $3.3\times 10^{-7}$  & 0.13\\
ESO3020120   & 3.04  & 3.19  & 1.65  & $5.1\times 10^{-8}$  & 0.01\\
ESO4250180   & 5.12  & 7.19  & 0.92 	& $8.1\times 10^{-8}$  & 0.12\\
ESO4880049   & 3.44  & 2.53  & 3.37 	& $1.3\times 10^{-7}$  & 0.06\\
ESO3050090   & 2.2  & 3.14  & 0.93  	& $1.6\times 10^{-8}$  & 0.06\\
U11454       & 5.18  & 3.32  & 4.42 		&$4.0\times 10^{-7}$  & 0.81\\
U11648       & 4.46  & 3.53 & 2.88  		& $1.9\times 10^{-7}$  & 5.32\\
U11819       & 5.82  & 4.63  & 2.87  		& $3.24\times 10^{-7}$  & 0.21\\
U5750        & 3.23  & 7.75  & 0.31  		& $1.1\times 10^{-8}$ & 0.01\\
U11748       & 7.94  & 1.07  & 99.33  	& $2.0\times 10^{-5}$  & 3.67\\
U11616       &  4.59 & 2.53  & 5.96  		& $4.2\times 10^{-7}$  & 0.20\\
U6614        & 6.19  & 2.85  & 8.54  		& $1.1\times 10^{-6}$ & 2.35\\
\hline \hline
\end{tabular}
\caption{Best fit values for $(a,\beta_0)$ for several low surface brightness galaxies. Those values can be 
translated into values for the central density and pressure $ (\rho_0,p_0) $. We
have taken $R_\star=1~$Kpc, $M_\star=10^{10} M_\odot$, 
so that $q=5 \,\times 10^{-7}$. See bellow. }\label{table1}
\end{table}
\begin{figure*}
\includegraphics[angle=0,width=1.\textwidth]{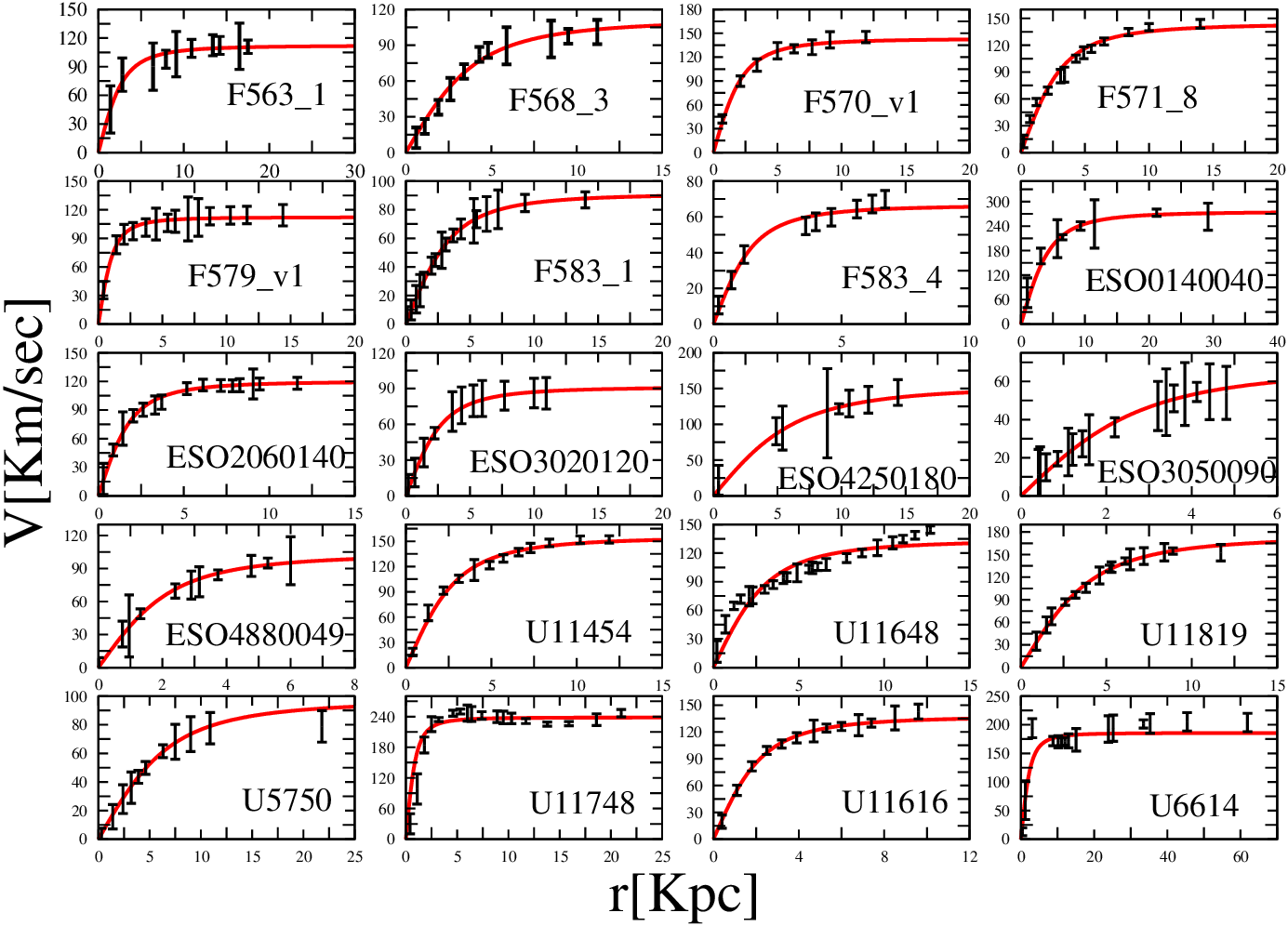}
\caption{Rotational Curve data from 20 Low Surface Brightness galaxies from
de Blok \& Bosma (2002). Red lines correspond to the best fit curve using PSS's rotational curve profile.}\label{Fig:RCdata}
\end{figure*}

Hence, it is possible to re-write the mass, the density and the pressure for a
halo satisfying the PSS's rotational velocity profile, in terms of the central pressure and
density. Rewriting the equation of state, Eq.(\ref{eq:equation of state1}), in this way we obtain:
\begin{equation}
\bar{p}(\bar{\rho})=\bar{p}_0\left(\frac{3}{4}\left(\frac{\bar{\rho}}{\bar{\rho}_0}\right) -
\frac{1}{16}\left(1 - \sqrt{1 + 24\left(\frac{\bar{\rho}}{\bar{\rho}_0}\right)}\right)\right)\,.\label{eq:equation of state2}
\end{equation}

Some comments about this equation are in order. The equation of state Eq.~(\ref{eq:equation of state2})
reduces to the barotropic equation of state $\bar p=\frac{3 \bar p_0}{4\,\bar{\rho_0}}\,\bar{\rho}$ in the 
limit where $\bar{\rho} \ll \bar{\rho}_0$. This has a natural explanation: $\bar{\rho}$ is a decreasing 
function, with maximum value 
$\bar{\rho}_0$ at $x=0$. In the limit $x \to \infty$ the density decreases and the rotation curve profile 
tends to a constant value, that is, the behavior of an ideal gas which has a barotropic equation of state.  

Eqs.~(\ref{eq:equation of state1}) and (\ref{eq:equation of state2}) have two 
free parameters;  either $\beta_0$ and $a$ or $\bar{\rho}_0$ and 
$\bar{p}_0$. In order to determine the values of those free parameters and advance 
with our method, we will use observational data of the rotational curves of several low surface brightness
galaxies, reported in de Blok \& Bosma (2002).
The procedure for a given galaxy, consists in determining the best values of  
$\beta_0$ and $a$ that fits the observed rotational curve through a $\chi^2$ analysis.

Let us for instance use the data for the galaxy ESO3050090. The observed rotational velocity
as a function of the radial distance from the center of the galaxy is shown in  
Fig.~(\ref{Fig1}). The best fit points are obtained by minimizing the function
\begin{equation}
\chi^2=\sum_i \left(\frac{\beta_{theo}-\beta_{exp_i}}{\delta \beta_{exp_i}}\right)^2\,,\label{eq:chi}
\end{equation} 
where $i$ runs up to the number of points in the data and $\beta_{theo}$ is computed 
according to the velocity profile under consideration, i.e. Eq.~(\ref{eq:beta_sal}).
$\delta \beta_{exp_i}$ is the error in the determination of
the rotational velocity.

Note that the terminal velocity is of the order of $100 ~\mbox{Km/s}$, 
so that, using the Newtonian relation for the mass of the halo: $m(r)=r\,{v_t}^2/G$, taking 
$G=4.299\,\,10^{-6}\,\frac{{\rm Kpc}\,\left({\rm km/s}\right)^2}{M_{\rm Sun}}$ and the characteristic 
distance $R_\star=1~$Kpc; the characteristic mass of the system is $M_\star=10^{10} M_\odot$. 
With these values of $M_\star$ and $R_\star$, we obtain $q=5 \,\times 10^{-7}$ and the characteristic values 
of the density and the pressure are:
$\rho_\star=1.62\,10^{-22}\,\frac{\rm grm}{{\rm cm}^3}$ 
and $p_\star=1.46\,10^{-1}\,\frac{\rm grm}{{\rm cm}\,s^2}$. 
We will use these characteristic values from now on.

After the minimization of the $\chi^2$ function we can obtain confidence level regions
for the parameters $\bar{\rho}_0$ and $\bar{p}_0$ which are mapped directly to $\rho_0$ and
$p_0$ through the characteristic values $\rho_\star$ and $p_\star$.
Such regions are shown in the inner panel of Fig.~(\ref{Fig1}). It is shown the
allowed region of parameters at $68\%$ (green region) and $90\%$ (red region) confidence level
for the central density $\rho_0$ and central pressure $p_0$. 
Hence, we have a prediction of the central pressure and density for the galaxy
obtained from the rotational curve data only.

The same analysis can be performed for any galaxy. For definitiveness
 we will use the data shown in Fig.(\ref{Fig:RCdata}) taken from Blok \& Bosma (2002). 
\begin{figure}
\includegraphics[angle=0,width=0.49\textwidth,height=!,clip]{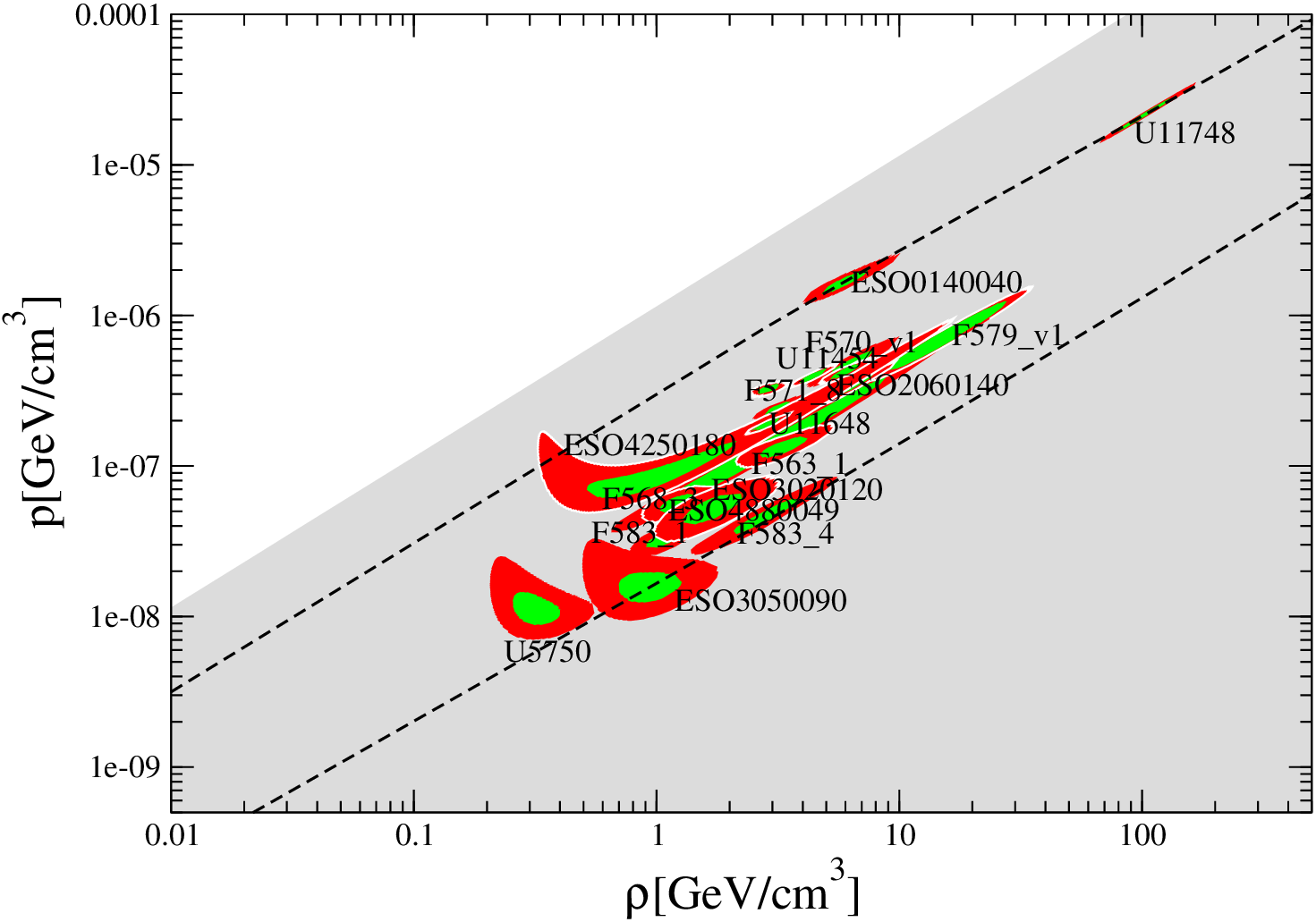}
\caption{Central density $\rho_0$ and pressure $p_0$ for all example galaxies fitted as expressed in Table \ref{table1}.
Green regions corresponds to $68\%$ C.L. and red regions corresponds to $90\%$ C.L. 
Dashed lines are the evaluation of the dark matter equation of state, Eq.~(\ref{eq:equation of state2}) 
for the best fit point  $(\rho_0,p_0)$ for U11748 (upper line) and ESO3050090 (lower line). 
Grey region is the allowed region obtained with cosmological data obtained by Muller (2005).}\label{Fig3}
\end{figure}

The best fit points for each galaxy are shown in Table~(\ref{table1}), either
for $(a,\beta_0)$ or $(\rho_0,p_0)$. We also present the goodness of the fit, which  is given by
the ratio of the minimum value of $\chi^2$ over the degrees of freedom, $\chi^2_{min}/\mbox{d.o.f}$. 

The allowed regions for $\rho_0$ and $p_0$ at $68\%$ and $90\%$ confidence level obtained for 
each  galaxies are presented in Fig.~(\ref{Fig3}).  

The equation of state $\bar{p}(\bar{\rho})$, Eq.~(\ref{eq:equation of state2}), 
gives the functional relation as soon as $\bar{\rho}_0$ and $\bar{p}_0$ are
given. In addition to the allowed regions of the central pressure and density, we
have included in Fig.~(\ref{Fig3}) in dashed lines the equations of state  $p(\rho)$
that are obtained by evaluating Eq.~(\ref{eq:equation of state2}) using the best fit values 
$\rho_0$ and $p_0$ 
obtained by minimizing the $\chi^2$ using the data of the galaxies U11748 and ESO3050090.
As can be seen, most of the allowed regions for $(\rho_0, p_0)$, and thus, their corresponding 
equation of state, for our sample of 
galaxies are contained between those two dashed lines.
In that sense, the velocity profile provides an unique equation of state for all galaxies, where the 
only free parameters are the central density $\rho_0$ and central pressure $p_0$.
From our limited sample of galaxies, we obtain:
\begin{eqnarray}  
0.93 ~\mbox{GeV/cm}^3<&\rho_0&<99 ~\mbox{GeV/cm}^3 \quad \mbox{and}\nonumber\\  
1.6\times 10^{-8}~\mbox{GeV/cm}^3<&p_0&<2\times 10^{-5}~\mbox{GeV/cm}^3\,,
\end{eqnarray} 
as allowed intervals.

In the next section we will discuss  how much those regions depend on the particular selection of 
the halo model.

Finally, let us comment about previous bounds on the equation of state obtained by analyzing cosmological
observations. In Fig.~(\ref{Fig3}) we have included on gray the limit found in Muller (2005). 
This limit was obtained assuming a barotropic equation of state for the dark matter $p=\omega_{DM}\rho$.
The allowed values for $\omega_{DM}$ are $-1.5\times 10^{-6}<\omega_{DM}<1.13\times 10^{-6}$. 
Similar analysis was done recently, including the latest results from Planck, 
Xu \& Chang (2013), it was obtained that 
$\omega_{DM} \in  7.07^{+7.46}_{-7.47}\times 10^{-4}$, which is less stringent than 
the limit obtained by Muller (2005).
In contrast, our results imply that  if dark matter is modeled as a perfect fluid with
a barotropic equation of state, it is not possible to reproduce the 
rotational curve velocity profiles. However, as can be seen in Fig.~(\ref{Fig3}),
the central pressure and densities of
all the galaxies we have analyzed are below the limit found in Muller (2005). Hence, 
we conclude that it could be interesting to repeat the analysis done 
for the cosmological cases but using the dark matter equation of state Eq.~(\ref{eq:equation of state2})
instead that the barotropic relation and to find out if there is consistency at galactic and cosmological scales
with an unique equation of state.

\section{Dark Matter equation of state from the mass density profile}\label{density_profiles}

Our derivation for the equation of state for the dark matter only needs the velocity profile
$\beta^2(x)$. For the cases when the mass density profile $\bar{\rho}(x)$ is given, it can be used
to determine $\beta$.
Indeed, given $\bar \rho$, the mass function $n(x)$ is computed from Eq.~(\ref{eq:Nrho})
and then, using Eq.~(\ref{eq:Nn}) $\beta(x)$
is obtained. Furthermore, the other metric coefficient $\Phi(x)$ is computed using Eq. (\ref{eq:phi_v}) and
finally, the pressure $\bar{p}(x)$ is obtained via integration of Eq.~(\ref{eq:Nt}).
In the following we will compute $n(x), \beta^2(x), \Phi(x)$ and $\bar{p}(x)$ for several 
mass density profiles. Then, the functional relationship $\bar{p}(\bar{\rho})$ will be derived.    
We will consider the Pseudo-isothermal, Einasto, Navarro-Frenk-White  and the Burkert dark
matter density profiles.

Note that to perform such procedure in the general relativistic case is more involved. In
this case, one obtains a system of equations for $\Phi'$ and $\bar{p}$ and, once it is solved,
the velocity profile $\beta^2(x)$ is determined by means of Eq.~(\ref{eq:phi_v}). In any case, as 
it is mentioned above and shown in the appendix,
the Newtonian results agree with the relativistic ones to high precision, so that it is
enough to consider the Newtonian case. 

\subsection{Pseudo isothermal mass density profile}\label{pseudoisoterma}

An interesting example of a mass density profile which is regular at the origin 
is given by the pseudo isothermal model, where the
dark matter density profile takes the form:
\begin{equation}
\bar{\rho}(x)=\frac{A}{x^2+x_c^2},
\end{equation}
where $A$ and $x_c$ are parameters used to adjust the model to the observations. 

In this case, the velocity profile will be given by
\begin{equation}
\beta^2(x)=3\,q\,A\,\left(1-\frac{\arctan\left(\frac{x}{x_c}\right)}{\frac{x}{x_c}}\right),
\end{equation}
and the geometric potentials are, the mass function:
\begin{equation}
n(x)=3\,A\,x\left(1 - \frac{\arctan\left(\frac{x}{x_c}\right)}{\frac{x}{x_c}}\right)\,,
\end{equation}
and the gravitational potential
\begin{equation}
\frac{\Phi(x)}{c^2}=3\,q\,A\,\left(\ln\,\sqrt{1+\left(\frac{x}{x_c}\right)^2} +
\frac{\arctan\left(\frac{x}{x_c}\right)}{\frac{x}{x_c}}\right).\label{potencialIso}
\end{equation}
From these expressions, by integration of  Eq.~(\ref{eq:Nt}),
one finally obtains the pressure of a pseudo isothermal halo:
\begin{equation}
\bar{p}(x)=\frac{3\,q\,A^2}{{2\,x_c}^2}\,\left(\frac{\pi^2}{4} -
2\,\frac{\arctan\left(\frac{x}{x_c}\right)}{\frac{x}{x_c}}
 - \left({\rm arctan}\left(\frac{x}{x_c}\right)\right)^2
\right).
\end{equation}

In the last expression, 
we have chosen the integration constant such that the pressure vanishes far from the center of the halo.

Identifying the values of the density and the pressure at the center of the halo, $\bar{\rho}_0$
and $\bar{p}_0$, we relate them with the parameters of the model by 
\begin{equation}
\bar{\rho}_0=\frac{A}{x_c^2}\quad \mbox{and}\quad \bar{p}_0=\frac{3\,q\,A^2}{8{x_c}^2}\,(\pi^2 - 8)\,,\label{eq:rho0p02}
\end{equation}
so that we can write down the previous expressions for the gravitational potential, the mass
function, the pressure and the density in terms of the 
central pressure and the central density. 
In particular,  the gravitational potential can be written as:
\begin{equation}
\frac{\Phi(x)}{c^2}=\frac{8\,\bar p_0\,}{\bar \rho_0 (\pi^2-8)}\left(\ln\,\sqrt{2-\left(\frac{\rho_0}{\rho}\right)^2} +
\frac{\arctan\sqrt{\frac{\bar{\rho}_0}{\bar{\rho}} - 1}}{\sqrt{\frac{\bar{\rho}_0}{\bar{\rho}} - 1}}\right)\,.
\end{equation}
%
square root remains always positive.

Finally, the equation of state for the Pseudo-Isothermal density profile in terms of the central parameters is:
\begin{eqnarray}
&&\bar{p}(\bar{\rho})=\frac{8\bar p_0}{\pi^2-8}\,\times\\
&&\left[\frac{\pi^2}{8} - 
\frac{{\rm arctan}\sqrt{\frac{\bar{\rho}_0}{\bar{\rho}} - 1}}{\sqrt{\frac{\bar{\rho}_0}{\bar{\rho}} - 1}} - 
\frac12\,\left({\rm arctan}\sqrt{\frac{\bar{\rho}_0}{\bar{\rho}}  - 1}\right)^2\right]\,.\nonumber
\label{eq:pvsrho_iso_t}
\end{eqnarray}

Although it is not a linear relation between the pressure and the density, 
when the density is much smaller than the central density $\rho \ll \rho_0$,
the equation of state has the limit

\begin{equation}
\bar p(\bar{\rho}\ll\bar{\rho_0})=\frac{4 \bar{p}_0}{\bar{\rho}_0\,\left(\pi^2 - 8\right)}\,\bar{\rho}\,,
\label{eq:edo-pse-iso-fl}
\end{equation}
which is a barotropic equation of state,  very similar to 
the equation of state derived in section \ref{ajustes} in the same limiting case.

It is possible to fit the data from the observations of the set of galaxies as done in the previous
section. 

\begin{figure}
\includegraphics[angle=0,width=0.46\textwidth,height=!,clip]{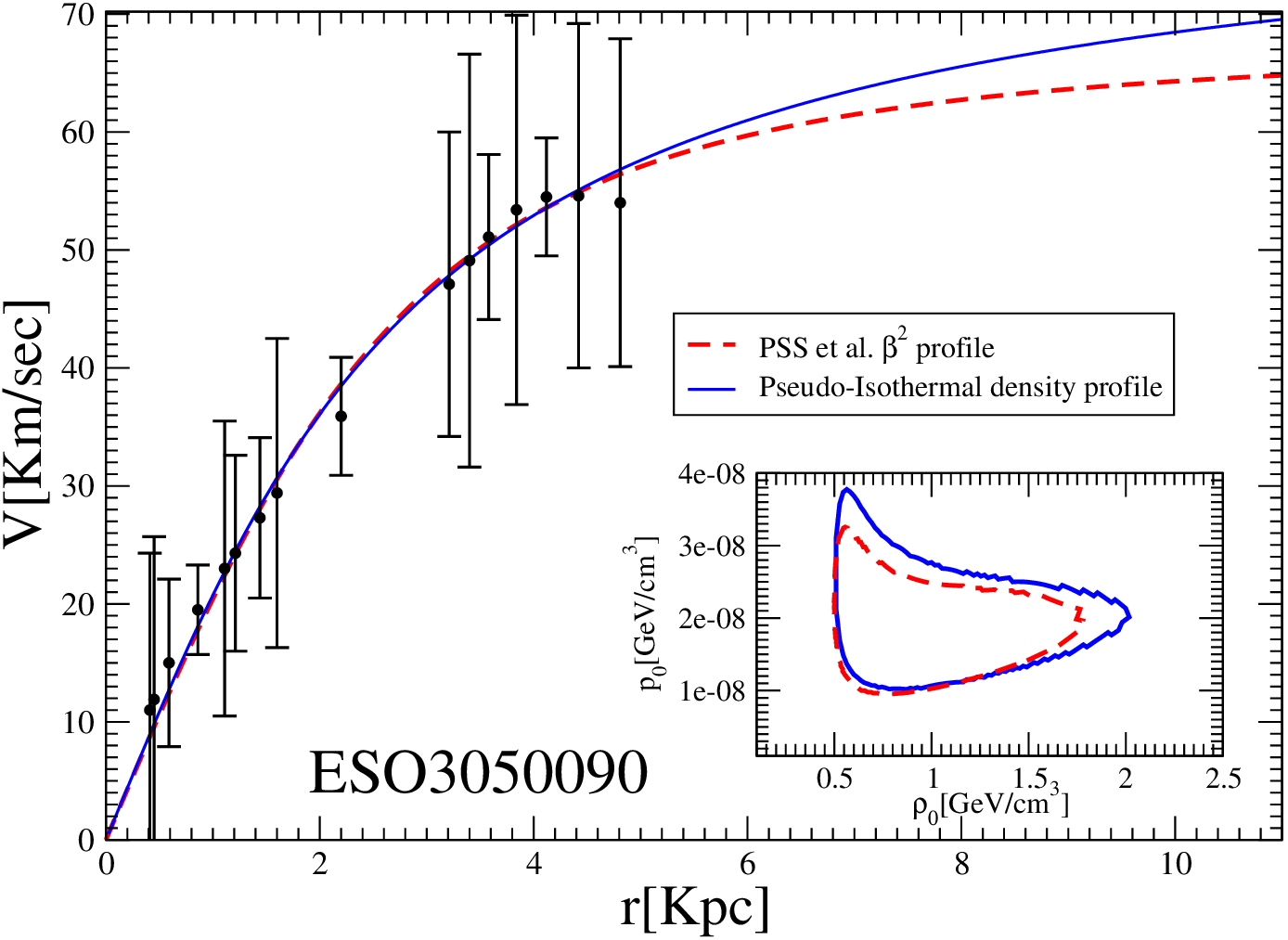}
\caption{Data points of the rotational velocity for the galaxy ESO3050090 and 
best fit curve using PSS's velocity profile (dashed line) and velocity profile obtained  with the
Pseudo-isothermal dark matter density (solid line).
Inner figure: Different predictions for the central density and pressure for both velocities profiles}\label{Fig4}
\end{figure}
\begin{figure}
\includegraphics[angle=0,width=0.46\textwidth,height=!,clip]{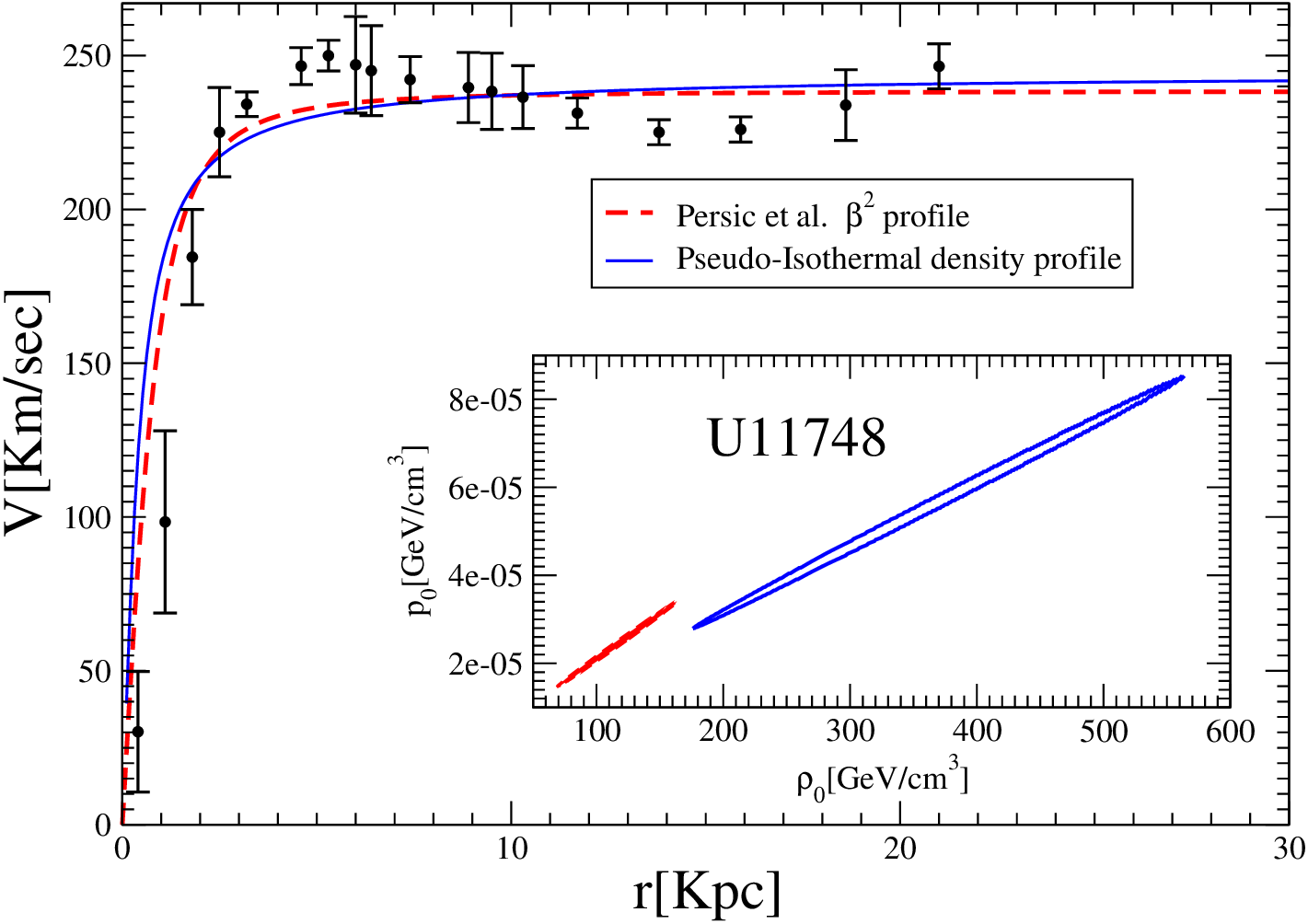}
\caption{Data points of the rotational velocity for the galaxy U11748 and 
best fit curve using PSS's velocity profile (dashed line) and velocity profile obtained  with the
Pseudo-isothermal dark matter density (solid line).
Inner figure: Different predictions for the central density and pressure for both velocities profiles}\label{Fig5}
\end{figure}

The pressure as a function of the radial distance $p(x)$ obtained by using either the 
PSS velocity profile $\beta^2(x)$ or the Pseudo-Isothermal mass density profile $\bar{\rho}(x)$ 
is regular at $x=0$. 
This allows us to rewrite the velocity profile's free 
parameters in terms of the central density and the central pressure $(\rho_0,p_0)$ with the 
help of Eq.~(\ref{eq:rho0p02}) as done in previous section.

Now we address the question of how much the predictions for $\rho_0$ and $p_0$ depend on 
the election of either the velocity or the mass density profiles, at least for the 
two profiles we have considered that have regular behavior at $x=0$.
Let us consider again the data for the rotational curve of the galaxy ESO3050090.  

As it is shown in Fig.~(\ref{Fig4}), the data can be nicely fitted with either the
PSS's velocity profile or the velocity profile derived from the  Pseudo-isothermal density profile.
The allowed regions for the $\rho_0$ and $p_0$ at $90\%$ confidence level are shown in the
inner figure. As can be observed, both regions coincide independently of the halo model
used. From this figure we may infer that the prediction for the central density and pressure
is independent of the parametrization used to fit the rotational curve data.

Nevertheless, the same procedure can be done for the data of the Galaxy U11748. The fits for
the rotational velocity data are shown in Fig.~(\ref{Fig5}). Again there is a 
set of $(\rho_0,p_0)$ that can give a nice fit to the data, but in contrast to the case
of the galaxy ESO3050090, the regions for $\rho_0$ and $p_0$ obtained in each case
are different depending on the functional 
parametrization of $\beta^2(x)$ that is used. This is shown in the inner figure of Fig.~(\ref{Fig5})
that shows the allowed regions for $\rho_0$ and $p_0$ at $90\%$ C.L. for each velocity profile.

A closer look at Fig.~(\ref{Fig4}-\ref{Fig5}) can help us to understand the reason why
in one case it seems that there is a unique determination of the central density and pressure while
in other case we found a different determination of the central pressure an density. 
The data points of the galaxy ESO3050090 exhibit large error bars,
hence the allowed values for $(\rho_0,p_0)$ spread over a large region.

On the other hand, the data points for U11748 have smaller error bars
and there are more data points that are located for either small or 
large values of $x$. The last data points are relevant since the flat behavior of the rotational curve
is present. Thus, the fit should consider both regions: the inner region and the flat behavior and
the minimization of $\chi^2$ will found a more restricted region for $(\rho_0,p_0)$ that produces 
a rotational curve consistent with all data points. Hence, we may conclude that
the prediction of the central density and pressure will depend on the quality of the velocity profile we are using.
It is desirable to have data points with small errors at a wide range of distances from the center of the galaxies. 

The determination of the central density has a direct implication in experiments looking for direct or
indirect detection of dark matter. 
From our analysis in the galaxy U11748 we have found that the determination of the central density 
strongly depends on the density or velocity profile we are using. We will discuss more about this point in section 
\ref{discussion}.

We end this sub-section by presenting the corresponding results for the isothermal sphere. In this case, the
density profile is given by
\begin{equation}
\bar{\rho}_{\rm Iso}(x)=\frac{A}{x^2},
\end{equation}
and it is straightforward to determine the corresponding mass, velocity profile and the pressure,
from Eqs.~(\ref{eq:Nrho}-\ref{eq:Nt}):
\begin{eqnarray}
\bar{n}_{\rm Iso}(x)&=&3\,A\,x, \\
{\Phi'}_{\rm Iso}(x)/c^2&=&3\,q\,\frac{A}{x}, \\
{\beta^2}_{\rm Iso}(x)&=&3\,q\,A, \label{betaiso}\\
\bar{p}_{\rm Iso}(x)&=&\frac{3\,q\,A^2}{2\,x^2},
\end{eqnarray}
where we have set to zero the integration constant in the pressure. In this way, we obtain that
the equation of state for the isothermal dark fluid sphere is
\begin{equation}
\bar{p}_{\rm Iso}=\frac{3\,q\,A}{2}\,\bar{\rho}_{\rm Iso}.
\end{equation}
It is a barotropic equation of state and actually coincides, as it should,
with the limit of the equation of state for the pseudo isothermal halo far from the central region, 
Eq.~(\ref{eq:edo-pse-iso-fl}) when $\bar p_0,\bar \rho_0$ are rewritten in terms of the original parameters 
Eq. (\ref{eq:rho0p02}).

In this case, ${\beta^2}_{\rm Iso}(x)$ is a constant, and could play the role of a measure of
an analogous to the temperature of the dark halo. However, it is impossible to fit 
the rotational velocity curve in this case and then,  a isothermal equation of state
can not describe the dark matter if it is desired a good fit with the rotational velocity 
when dark matter is treated as a perfect fluid.

Nevertheless, in the regions where the rotational velocity profile is constant, the isothermal
sphere can give some hints on the dark matter properties. 

\subsection{Einasto's density profile}\label{Einasto}

The derivation of the equation of state can be performed for any given rotational velocity profile or
density profile proposed. Previously we have shown profiles from which the behavior of both the density and the pressure 
at the center are regular.  

Nevertheless, there are many other dark matter density profiles.
Among them, a three parametric function for the
density profile was proposed by Einasto (1965) several years ago and has recently received more attention.
It is given by 
\begin{equation}
\rho_{\rm Ein}=\rho_0\,e^{-\frac{2}{L}\left(\left(\frac{x}{x_c}\right)^L - 1\right)},\label{eq:rho_ein}
\end{equation}
where $\rho_0, x_c$ and $L$ are free parameters. This case is also regular at the origin, $\rho(x=0)=\rho_0\,e^\frac{2}{L}$,
for finite $\rho_0, L$, although it can be very cuspy for small values of the parameter $L$.

It is possible to obtain an analytic expression for the velocity profile by integration of Eq.~(\ref{eq:Nrho}) and with the help
of Eq.~(\ref{eq:Nn}) (obtained through Maple16):
\begin{eqnarray}
{\beta^2}_{\rm Ein}&=&\frac{q\,\rho_0\,L^\frac{3\left(L+1\right)}{L}\,x^2\,
e^{-\frac{\left(\frac{x}{x_c}\right)^L - 2}{L}}}{4\,\left(8^{(\frac{1}{L})}\right)\,
\left(\frac{x}{x_c}\right)^\frac{L+3}{2}\,\left(L+3\right)\,\left(2\,L+3\right)}\,\times\nonumber \\
&&\left(L\,\left(2^{\frac{3(1+L)}{2\,L}} + 2^{\frac{3+L}{2\,L}}\,\left(\frac{x}{x_c}\right)^{-L}\,
\left(L+3\right)\right)\,\times\right.\nonumber \\
&&\left.{\rm WhittakerM}\left(-\frac{L-3}{2\,L}, \frac{2\,L+3}{2\,L}, \frac2L\,
\left(\frac{x}{x_c}\right)^L\right) + 
\right. \nonumber \\
&&\left.  2^{\frac{3+L}{2\,L}}\,\left(\frac{x}{x_c}\right)^{-L}\,
\left(L^2 + 6\,L + 9 \right)\,\times\right.\label{beta_einasto}\\
&&\left.{\rm WhittakerM}\left(\frac{L+3}{2\,L}, \frac{2\,L+3}{2\,L}, 
\frac2L\,\left(\frac{x}{x_c}\right)^L\right) \right)\,. \nonumber
\end{eqnarray}
\begin{figure}
\includegraphics[width=0.49\textwidth]{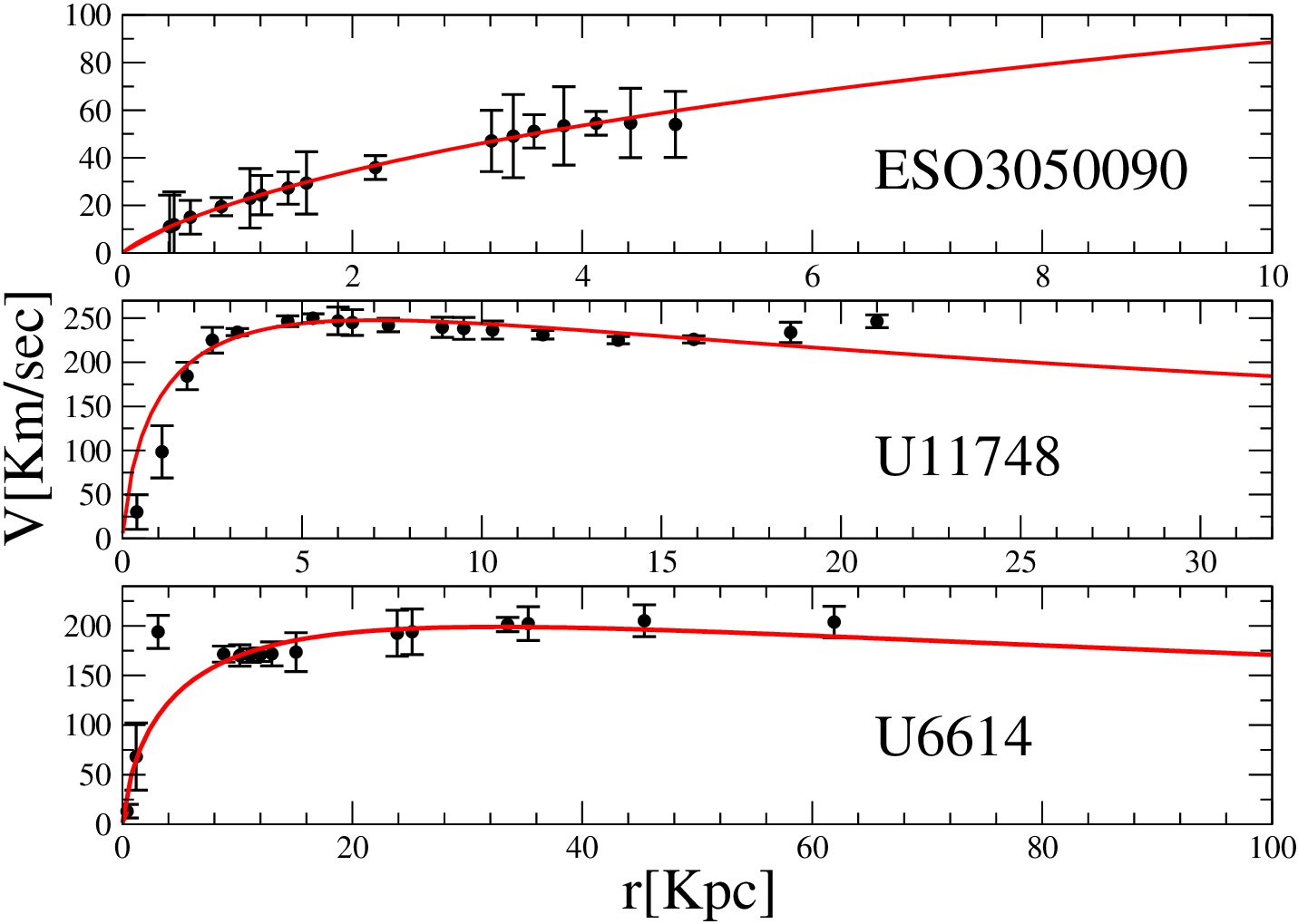}
\caption{Fit for some LSB galaxies by using the Einasto's velocity profile Eq.~(\ref{beta_einasto})}\label{Fig6}
\end{figure}

Using Eq. ~(\ref{beta_einasto}) it is possible to fit our set of velocities profiles. In 
Fig. ~(\ref{Fig6}) we give a couple of examples. 
Those galaxies where only data points for small values of the distance to the center of the galaxies are given 
can be easily fitted by the Einasto profile (for instance ESO3050090). But for those galaxies that
start to show the characteristic flat behavior for large values of $x$, it is more difficult to 
have a good fit. 
This behavior can be explained  through Eq.~(\ref{beta_einasto}) since
$\lim_{x \to \infty}\beta(x)_{Ein}=0$. In other words, Einasto's profile can not have an extremely large flat behavior for the
rotational curves.

\begin{figure}
\includegraphics[width=0.49\textwidth]{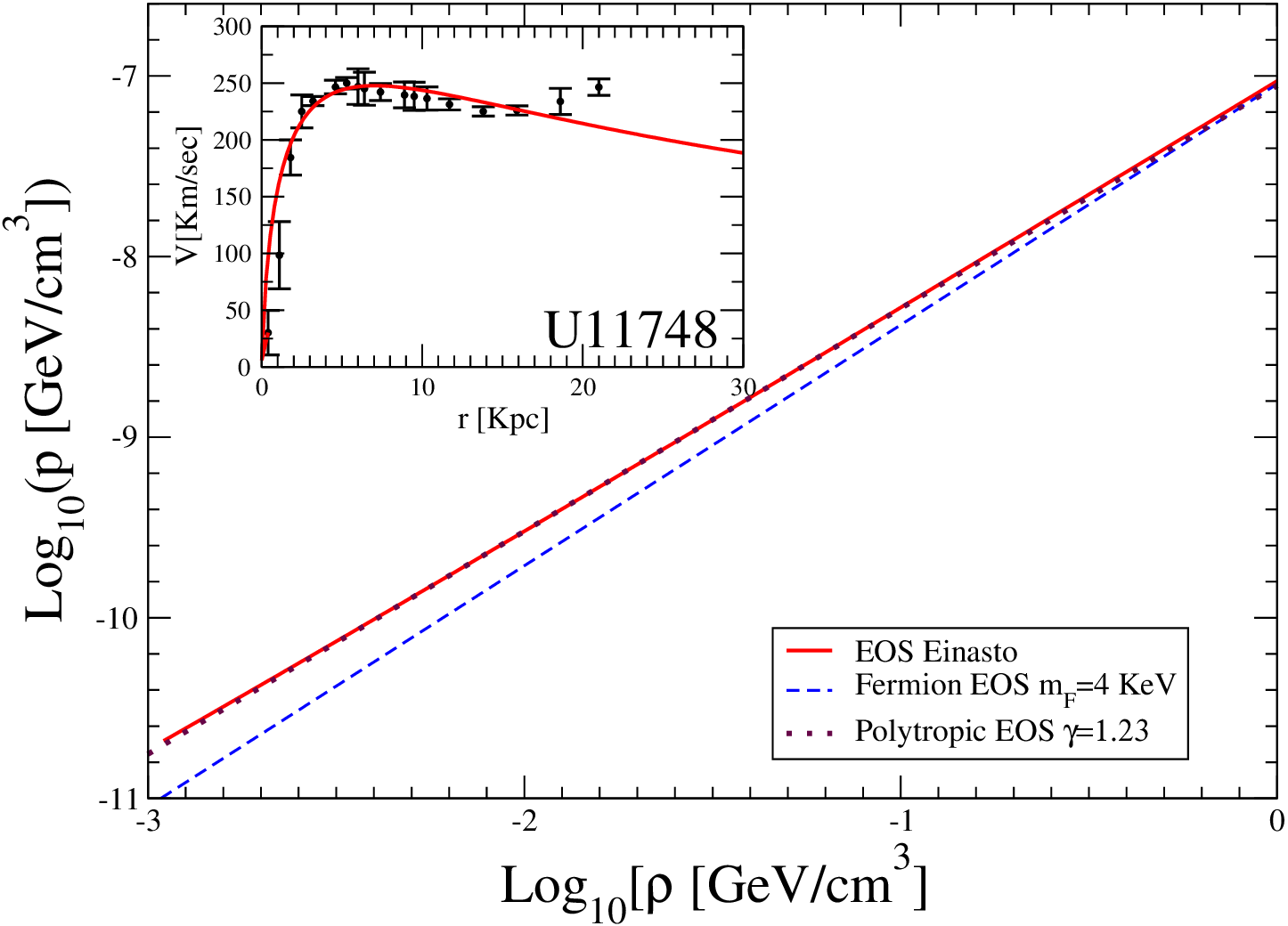}
\caption{Equation of state obtained using the Einasto's density profile. Dashed line shows the 
equation of state for a fermion with mass of $4$ KeV. }\label{Fig7}
\end{figure}

The mass function and the gradient of the gravitational potential 
are directly obtained from the velocity profile: $n_{\rm Ein}=\frac{x}{q}\,{\beta^2}_{\rm Ein}$, and
${\Phi'}_{\rm Ein}/c^2=\frac{{\beta^2}_{\rm Ein}}{x}$. Then, we can obtain the pressure by integration of Eq.~(\ref{eq:Nt}). 
We did not find analytical expression, but this can be done numerically. 
In Fig.~(\ref{Fig7}) we show the equation of state obtained with the bets fit parameters for the galaxy U11748.

Due to the form of $\beta(x)_{Ein}$ for large values of $x$, contrary to the PSS's velocity profile and
the Pseudo-isothermal density profile that show a barotropic equation of state for small values of the density,
the Einasto's equation of state have a different behavior. For instance, using the best fit values 
obtained with the data points of the galaxy U11748, the resulting equation of state 
have a polytropic behavior for small values of the density (in the limit $\rho \ll \rho_0)$.
That is, when $x \to \infty$ the Einasto equation of state behaves as $p \sim \rho^{\gamma}$ with $\gamma \simeq 1.23$.
(see Fig.~(\ref{Fig7})).

We will end this sub-section by comparing the Einasto equation of state with another polytropic equation that
arise when dark matter is assumed to be a free gas of fermions at zero temperature.
In the case when the equation of state is given, the study of an equilibrium configuration is described by the Tolman-Oppenheimer-Volkov, TOV, equations Tolman (1939) , Oppenheimer and Volkov (1939), see also Silbar and Reddy (2004). Notice that this is not the case described in the present work, as long as we derive the equation of state from the observations.

Indeed, the equation of state for non interacting fermions can be directly computed (see Narain et al. (2006) for instance):
\begin{eqnarray}
\rho&=&\frac{1}{\pi^2}\int_0^{k_F}k^2\sqrt{m_F^2+k^2}dk \nonumber\\
&=&\frac{m_f^4}{8\pi^2}\left((2z^3+z)(1+z^2)^{1/2}-\sinh^{-1}(z)\right)\,,\\
p&=&\frac{1}{3\pi^2}\int_0^{k_F}\frac{k^4}{\sqrt{m_F^2+k^2}}\nonumber\\
&=&\frac{m_f^4}{24\pi^2}\left((2z^3-3z)(1+z^2)^{1/2}+3\sinh^{-1}(z)\right)\,, \label{equation of state_fermion}
\end{eqnarray}
with $z=k_F/m_f$, $k_F$ the Fermi momentum and $m_F$ the mass of the hypothetical fermion
that will represent the dark matter particle.

Eqs.~(\ref{equation of state_fermion}) are in units where $\hbar=c=1$. The only free parameter is the mass
of the hypothetical dark fermion $m_F$. In Fig.~(\ref{Fig7}) in addition to the equation of state 
obtained by using the Einasto density profile we have added the resulting equation of state for a 
free fermion at zero temperature given by Eqs.~(\ref{equation of state_fermion}) for a fermion mass of $m_F=4~$KeV.
This comparison might suggest that the equation of state obtained by fitting either 
the rotational curve profiles or the dark matter density profile will not be mimicked by
the equation of state of a free fermion at zero temperature. In the low energy limit, the fermion has a 
polytropic equation of state with $\gamma=5/3$, which is a little bit far from the
index for the Einasto profile, or the PSS's velocity profile or the pseudo-isothermal, the last two
have a barotropic equation of state in the low energy limit. 

\subsection{Navarro-Frenk-White mass density profile}\label{NFW_profile}
\begin{figure}
\includegraphics[width=0.49\textwidth]{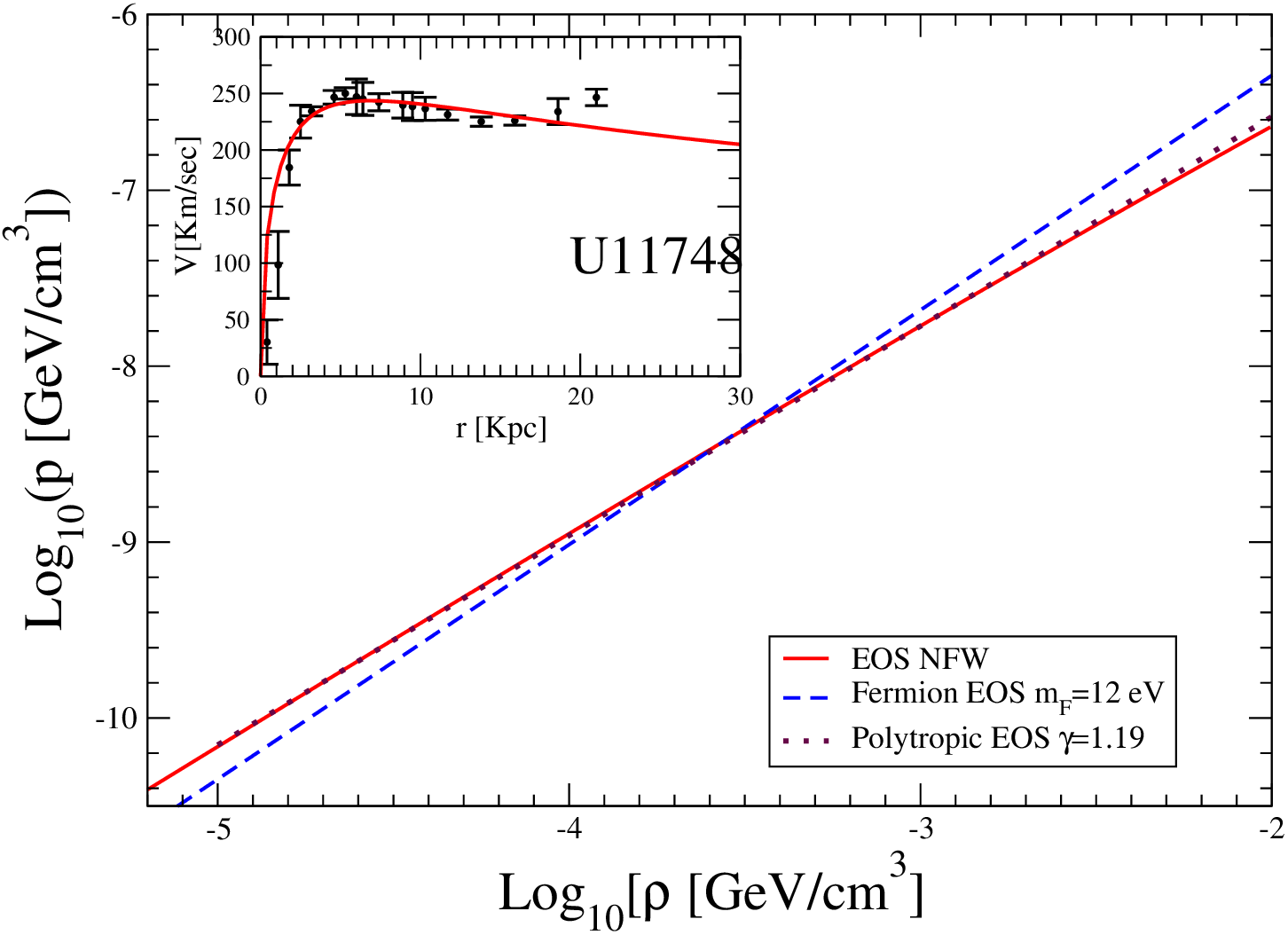}
\caption{Equation of state obtained using the NFW's density profile. 
Dashed line shows the equation of state for a fermion with mass 
of $4$ KeV. Dotted line corresponds to a polytropic fit with $\gamma=1.19$}\label{Fig8}
\end{figure}
Numerical simulations in the frame of the $\Lambda$-Cold Dark Matter paradigm predicts a cuspy dark matter
profile named the Navarro-Frenk-White mass density profile (NFW):
\begin{equation}
\bar{\rho}=\frac{\bar{\rho}_0}{x\,(x+x_c)^2}\,.
\end{equation}

As done in the previous section, it is straightforward to obtain 
the mass function, the velocity profile, and the gravitational potential from the NFW density profile, namely:
\begin{eqnarray}
n(x)_{\rm NFW}&=&3\bar{\rho}_0\,\left(\ln\left(1+\frac{x}{x_c}\right)-\frac{x}{x+x_c}\right)\,\\
{\beta^2(x)}_{\rm NFW}&=&\frac{3\,q\,\bar{\rho}_0}{x}\,\left(\ln\left(1+\frac{x}{x_c}\right)-\frac{x}{x+x_c}\right)\,,\label{eq:beta2NFW}\\
\Phi(x)_{\rm NFW}/c^2&=&-\frac{3 q \bar{\rho}_0}{x}\,\ln\left(1+\frac{x}{x_c}\right)\,,
\end{eqnarray}
where the integration constant in the mass function, $n$, has been chosen such that $n(0)=0$.

With these expressions, it is possible to obtain the pressure for the NFW which is given by:
\begin{eqnarray}
&&\bar p(x)_{NFW}=\frac{3\,q\,\rho_0^2}{2 {x_c}^4}\left[
3\,\ln\left(1+\frac{x}{x_c}\right)^2 - \,\ln\left(\frac{x}{x_c}\right) + \right. \nonumber \\
&& \left. \frac{x^3 - 5\,x_c\,x^2 - 3\,{x_c}^2\,x + {x_c}^3}{x^2\,\left(x+x_c\right)}\,\ln\left(1+\frac{x}{x_c}\right) 
+  \right. \nonumber \\
&& \left. + 6\,{\rm dilog}\left(1+\frac{x}{x_c}\right) - \ln\left(\frac{x}{x_c}\right) + \pi^2   +\right. \nonumber \\
&& \left. - \frac{x_c\,\left(7\,x^2 + 9\,x_c\,x + {x_c}^2\right)}{x\,\left(x + x_c\right)^2} \right]\,,
\end{eqnarray}
where ${\rm dilog}(x)=\int\limits_1^x\,\frac{\ln(t)}{1-t}\,dt$ is the dilogarithm function, and we have chosen
the integration constant such that the pressure vanishes at infinity.
It can be seen that the pressure at $x=0$ is not defined, hence, it is not possible to relate the free parameter $x_c$ with 
the central pressure. 

Nevertheless, with the help of Eq.~(\ref{eq:beta2NFW}) we can fit the rotational curves data by varying 
$\rho_0$ and $x_c$ and evaluate the density and the pressure for NFW and derive the corresponding equation of state.
A numerical evaluation of the resulting equation of state is shown in Fig.~(\ref{Fig8}).

The NFW profile implies, as the Einasto profile, a velocity profile that can not reproduce flat rotation curves for large
values of the radial distance. As can be seen from Eq.~(\ref{eq:beta2NFW}), the velocity tends to zero in the limit $x\to\infty$.
In the same way, the resulting equation of state will not present the barotropic behavior for small values of the density.
As can be seen in Fig.~(\ref{Fig8}), the NFW profile produces a polytropic 
equation of state with a value of $\gamma \simeq 1.19$,
which differs to the Einasto's case. 
For completeness we have also plotted the equation of state for a non 
interacting fermion, as we did for the Einasto profile. In both
cases we have used the best fit point for $x_c,\rho_0$ obtained from the fit to the 
galaxy U11748. Similar plots are obtained for other
galaxies from our sample listed in Table~(\ref{table1}). 

We end this sub-section by mentioning that an study along these lines was performed 
by Matos \& N\'u\~nez \& Sussman (2004), working in a Post-Newtonian description
of the NFW dark halo. An equation of state for the dark fluid was proposed, not derived as it is done in this work, considering a polytropic relation between the pressure and the density, which now we see, it was a good guess.

\subsection{Burkert mass density profile}\label{Burkert}

Finally, another interesting density dark matter profile was proposed by Burkert (1995).
\begin{equation}
\rho_{\mbox{Burkert}}=\frac{\rho_0}{\left(1+ \frac{x}{x_c}\right)\,\left(1
+ \left(\frac{x}{x_c}\right)^2\right)}\label{eq:rho_Burkert}
\end{equation}

As done in previous subsections, it is possible to obtain the mass function, the velocity profile and 
the gravitational potential for the NFW density profile, namely: 
\begin{eqnarray}
&&n_{\mbox{B}}(x)=\frac{3\,\rho_0\,{x_c}^3}{2}\,\left[\frac12\,
\ln\left(\left(1+\frac{x}{x_c}\right)^2\,\left(1 + \left(\frac{x}{x_c}\right)^2\right)\right) + \right. \nonumber \\
&& \left. -\mbox{tan}^{-1}\left(\frac{x}{x_c}\right)\right]\nonumber \\
&& \\
&&{\beta^2}_{\mbox{B}}=\frac{3\,q\,\,\rho_0\,{x_c}^3}{2\,x}\,\left[\frac12\,
\ln\left(\left(1+\frac{x}{x_c}\right)^2\,\left(1 + \left(\frac{x}{x_c}\right)^2\right)\right) + \right. \nonumber \\
&& \left.
-\mbox{tan}^{-1}\left(\frac{x}{x_c}\right)\right]\nonumber \\
&& \\
&&\frac{\Phi_{\mbox{B}}}{c^2}=-\frac{3\,q\rho_0\,{x_c}^2}{4\,x}\left[- \left(x-x_c\right)\,
\ln\left(1 + \left(\frac{x}{x_c}\right)^2\right) + \right. \nonumber \\
&& \left. + 2\,\left(x + x_c\right)\,\left(\ln\left(1 + \frac{x}{x_c}\right) -
{\mbox{tan}}^{-1}\left(\frac{x}{x_c}\right) \right) \right].
\end{eqnarray}

The pressure can be computed and an analytical expression can be found, with Wolfram's Mathematica, however
it is quite long and not enlightening, so we do not write it down, but use it for the numerical
evaluations.

Note that  $\lim_{x\to \infty}\beta(x)=0$, and although the decay of the rotational curve can be very small, at the end it is 
not possible to have flat rotation curves for extremely large values of $x$.
Furthermore, for low densities $\rho \ll \rho_0$, the resulting equation of state also has a polytropic behavior $p \sim \rho^\gamma$, very similar to the NFW profile, i.e. $\gamma=1.19$.

\section{Discussion}\label{discussion}
\label{sec:discussion}
\begin{table*}
\caption{Behavior of the density $\rho(x)$, pressure $p(x)$ and rotational velocity $\beta(x)$ for different
halo models at values of interest of the radial coordinate $x$.
}\label{Table2}

\begin{tabular}{|c|c|c|c|c|}
\hline
                            & Behavior $\rho(x=0)$ & Behavior $p(x=0)$ & $\lim_{x\to \infty}\beta(x)=v_\infty/c$ ? & Behavior $P(\rho \ll 1)$ ?\\
\hline
PSS's $\beta(x)$ profile & Regular              & Regular           & Yes                             & Barotropic                \\
Pseudo-isothermal DM profile& Regular              & Regular           & Yes                             & Barotropic                \\
Einasto DM profile          & Regular              & Not defined       & No                              & Polytropic $(\gamma \simeq 5/4)$\\
NFW DM profile              & Not defined          & Not defined       & No                              &  Polytropic $(\gamma \simeq 1.19)$\\
Burkert DM profile          & Regular              & Not defined       & No                              &  Polytropic $(\gamma \simeq 1.19)$\\
\hline
&&&&\\
\hline 
\hline
\end{tabular}
\end{table*}

The most direct evidence at galactic scale of the existence of dark matter comes from the rotational curves of galaxies. 
We have shown that by modeling the dark matter as a perfect fluid, it is possible, using only the data coming
from the rotational curves, to obtain the density and pressure of dark matter fluid, and the two metric 
functions (the mass and the 
gravitational potential). In particular, the 
velocity profile proposed by  Persic \& Salucci \& Stel (1996) and the Pseudo-isothermal dark matter 
density profile are two examples 
where all those functions are regular at the center of the galaxy. 
Even more, in some cases, an analytical functional relation of the pressure as a function of 
the density $p(\rho)$, that is, the equation of state for the dark matter, can be derived. 

The free parameters in the equation of state are fixed by fitting the observed rotational curve velocities 
with the velocity derived from each dark matter profile.

In Table~(\ref{Table2}) we have summarized the behavior and  some of the properties of the  density, 
the pressure and the velocity profile for the different halo models at two values of the radial coordinate. Furthermore,  the behavior 
of the equation of state for very low values of the density is also described for the velocity profile for all  dark matter 
density profiles used in this work. We want to stress that, contrary to other approaches,
where an equation of state for the dark matter fluid has been proposed, here we have 
derived such function by using the data from the rotational curves of a set of low surface brightness galaxies.
The resulting equation of state is neither barotropic or polytropic near the central part of the galaxy, as usually 
assumed, and we may conclude that
a barotropic equation of state can not fit the observed rotational curves of the galaxies.

\begin{figure}
\includegraphics[angle=0,width=.49\textwidth]{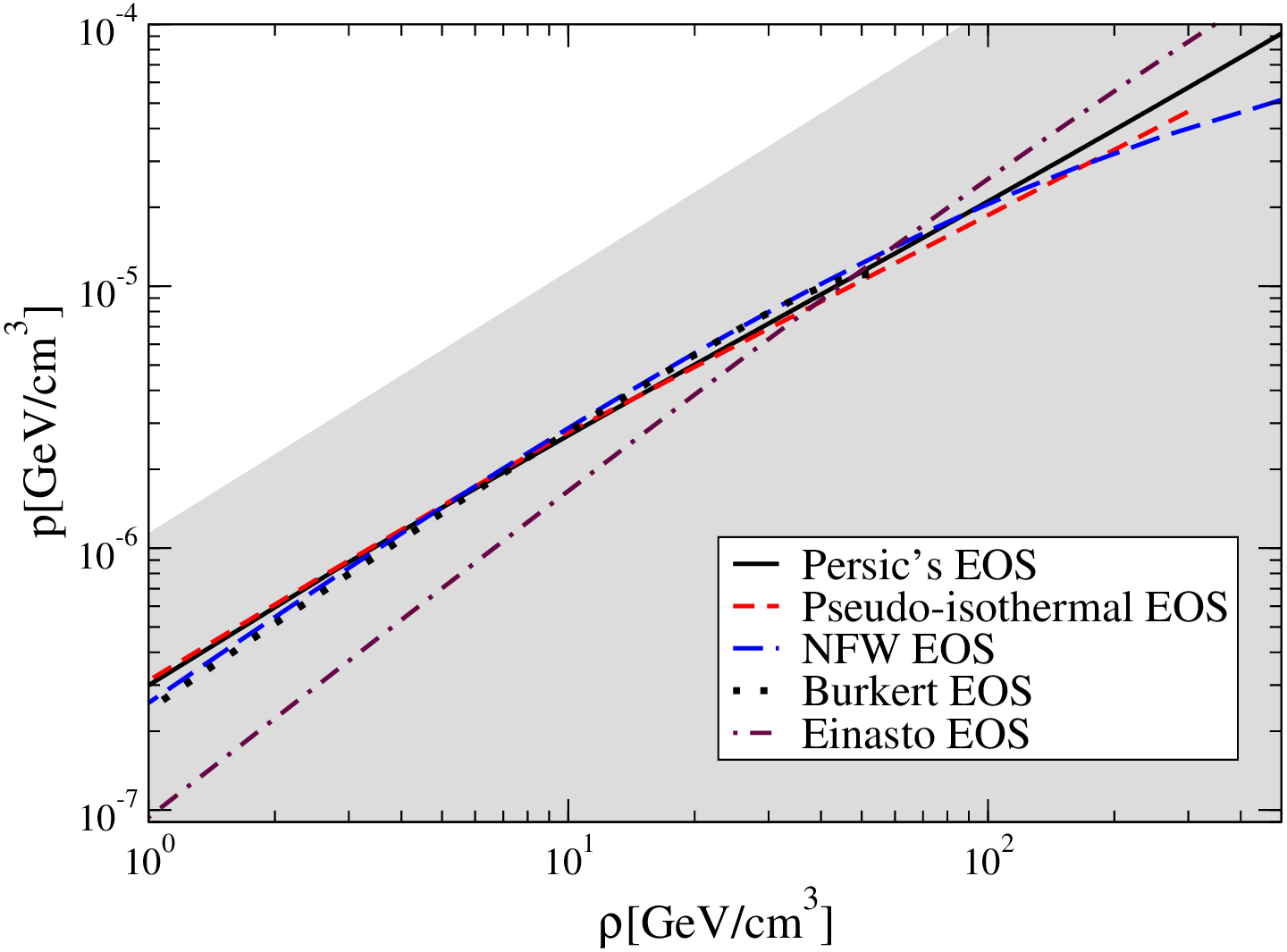}
\caption{All equations of state for the best fit parameter for the galaxy U11748.}\label{Fig9}
\end{figure}

Since both the pressure $p(x)$ and dark matter density profiles in the 
PSS's and Pseudo-isothermal profiles 
are regular at the origin, it is possible to related the central pressure $p_0$ and density $\rho_0$ with 
the only two free parameters of $\beta(x)$. This is important because allows us to understand 
that the equation of state has a functional 
dependence universal for all galaxies, and the free parameters, that is, the central pressure 
and density are related 
with the evolution history of the galaxy. Moreover, those equation of state has a barotropic 
limit for $\rho \ll \rho_0$ and 
flat rotation curves for large values of the radial coordinate.

Another interesting point, is that even if the dark matter density profile is regular at the origin, it does not
imply that the pressure will be regular. See for instance the Burkert and Einasto density profiles that has regular density 
but an ill-defined pressure at the center of the galaxy.
Furthermore, the Einasto, Burkert and Navarro Frenk White density profiles can not produce flat rotational curves 
for large values of the radial coordinate, and the resulting equation of state for low values of the density ($\rho \ll \rho_0$)
have a polytropic equation of state $p(\rho) \sim \rho^\gamma$ with $\gamma$ given in Table (\ref{Table2}).

We show in Fig.~(\ref{Fig9}) the equations of state obtained for all dark matter and 
velocity profiles
mentioned above. For definitiveness we have selected the galaxy U11748 and fitted the free parameters 
of each profile with its
rotational velocities data. Although any other galaxy can be used with similar results.
As can bee seen, there is a region where almost all equation of state superpose, and there are differences 
either for very low or very high 
densities. That is, in order to discriminate which equation of state is the most appropriate to describe 
dark matter, it will be necessary 
to study what characteristic compact objects produce the specific equation of state (i.e. the high density regime) or to study the
ultra low density regime (i.e. the intergalactic medium).

\begin{figure}
\includegraphics[angle=0,width=.49\textwidth]{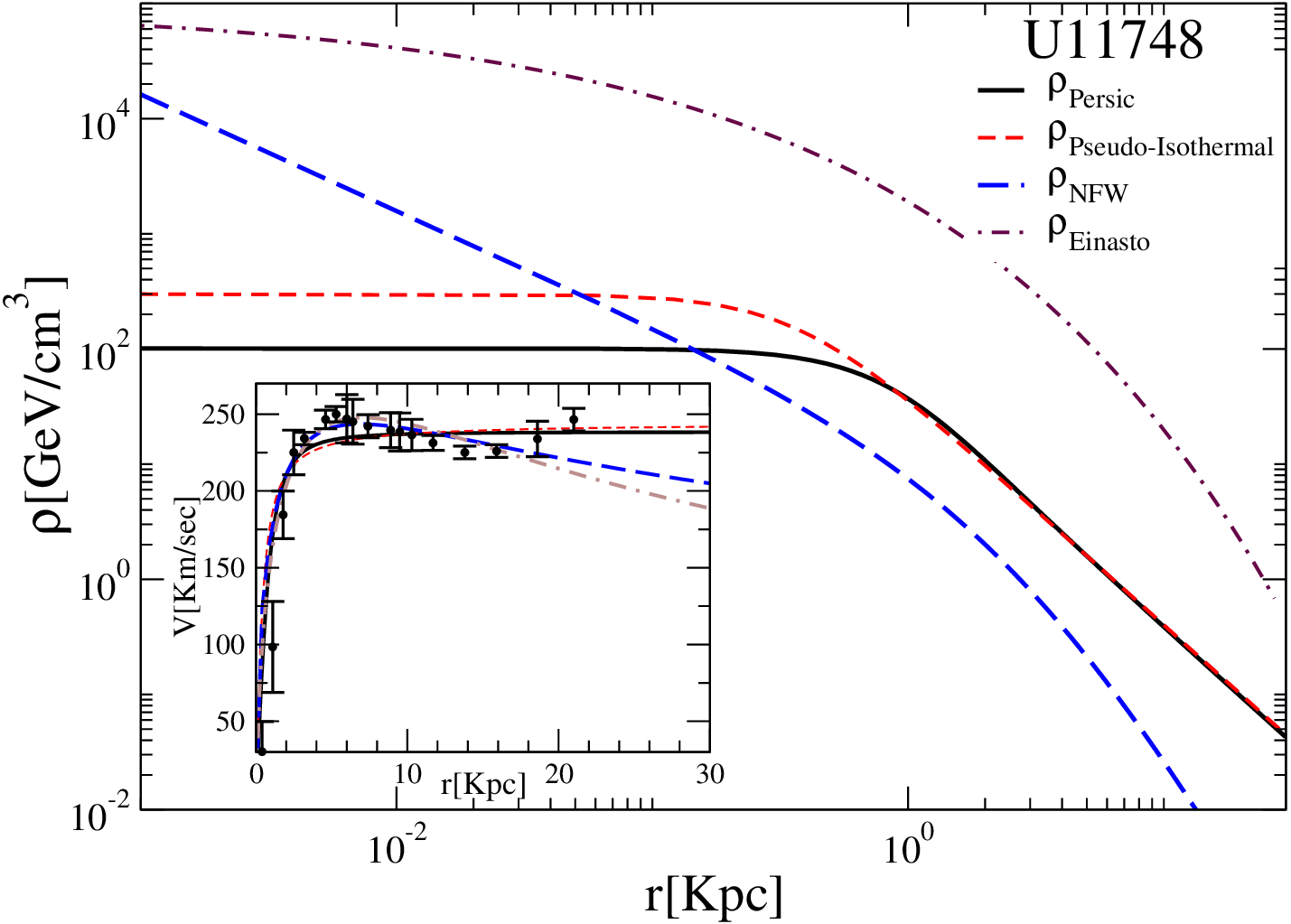}
\caption{Density profiles for the galaxy U11748 with the best fit points. Observe the difference 
in the determination of the local dark matter density}\label{Fig10}
\end{figure}

In the same manner as the behavior of the resulting equation of state in the high density regime depends on the 
dark matter density profile adopted, all other functions depend on the profile used.
This may have implications for the indirect detection of dark matter. See for instance Fig.~(\ref{Fig10}),
where
we have plotted the density as a function of the radial coordinate using the best fit values of
 our particular example of the U11748 galaxy.
As can be noted, the prediction for the density may change in orders of magnitude. Specially at the center of the galaxy. 
Indirect detection of dark matter depends strongly on the amount of dark matter at the center 
of the galaxy, hence the prediction may change drastically depending on the dark matter density profile used. 

Another quantity that may differ drastically depending on the profile used is the {\it Sound Speed} of the fluid.
Indeed, it is straightforward to compute the sound speed of the dark matter fluid $C_s^2=\partial P/\partial \rho$.
For the PSS's and Pseudo-isothermal profiles, analytical expressions can be obtained. For the PSS's velocity
profile:
\begin{equation}
C_s^2=\frac{3 p_0}{4 \rho_0}\left(1+\left(1+24\frac{\rho}{\rho_0}\right)^{-1/2}\right)\,,
\end{equation}
while for the Pseudo-isothermal profile
\begin{equation}
C_s^2=\frac{4p_0}{(\pi^2-8)(\rho_0-\rho)}\left(1-
\frac{\tan^{-1} \sqrt{\frac{\rho_0}{\rho}-1}}{\sqrt{\frac{\rho_0}{\rho}-1}}\right)
\end{equation}
\begin{figure}
\includegraphics[angle=0,width=.49\textwidth]{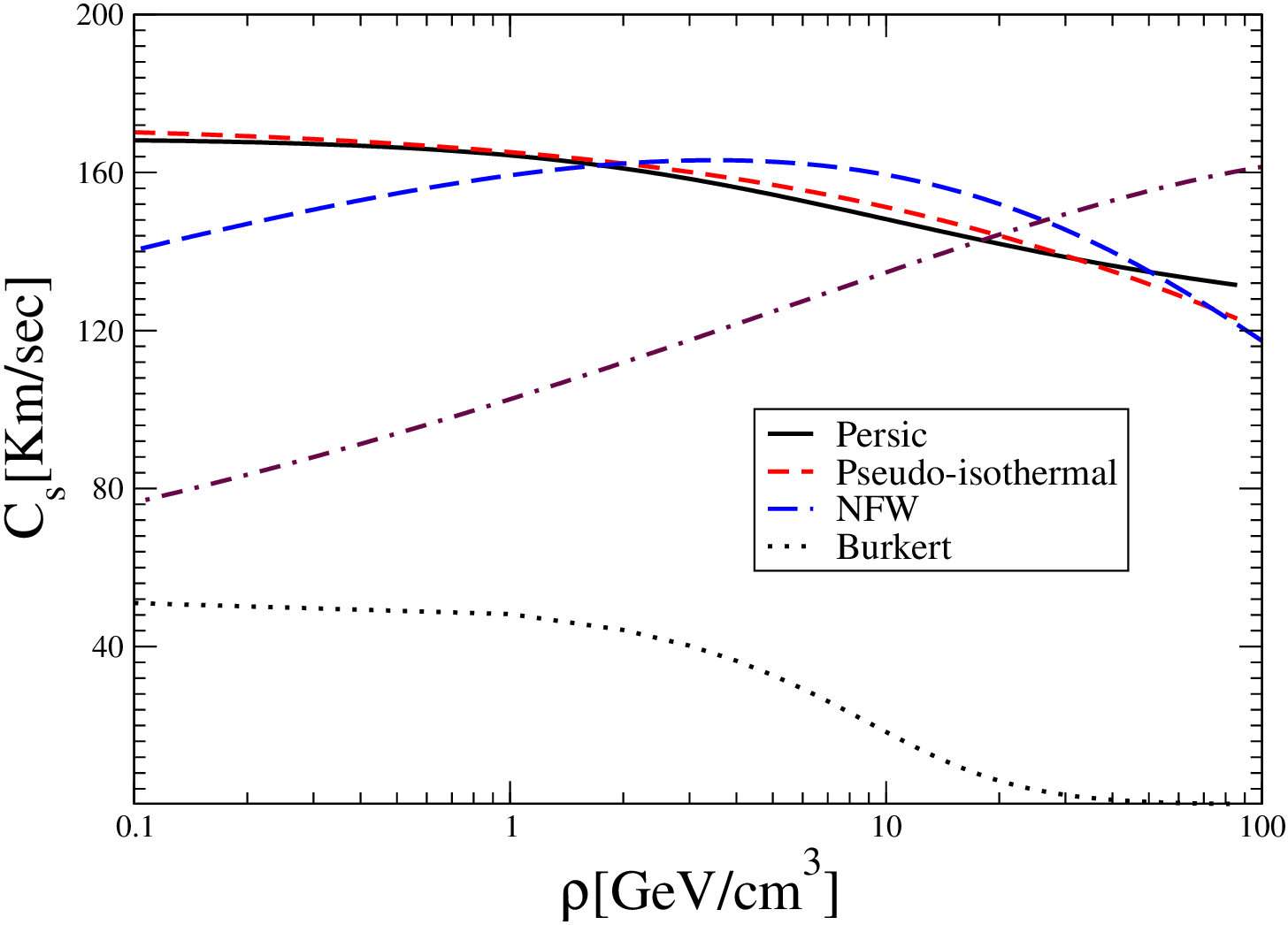}
\caption{The sound speed for the Dark matter fluid for different dark matter profiles. For definitiveness 
we have used the best fit points of
the galaxy U11748.}\label{Fig11}
\end{figure}

The NFW, Einasto and Burkert sound speed can be obtained by numerical derivation of the corresponding equation of state.
As can be see in Fig.~(\ref{Fig11}), the behaviors are very different, some of them change drastically as a 
function of the density.


This work was supported in part by DGAPA-UNAM grants IN115311, and IN103514, as well as a SNI-M\'exico
grant and Conacyt 167335.

\appendix

\section{General relativistic description}\label{apendice}
\label{App}

From the general relativistic equations, 
Eqs.~(\ref{eq:nRG}-\ref{eq:edoRG}), it can be obtained an equation (with no approximations) for 
the mass function as the only free function, as was done in N\'u\~nez et al. (2010). In this case, however, we are
more interested in the dark fluid properties, and, after some manipulation of these equations, 
we obtain an equation involving only the pressure, $\bar{p}$:
\begin{equation}
\bar{p}'+S(x)\,\bar{p} = T(x), \label{eq:pf}
\end{equation}
with 
\begin{eqnarray}
S(x)&=&-\frac{2\,\beta^2\,\left(1+\beta^2 - 2\,\beta^4 - x\,{\beta^2}'\right)}
{\left(1+\beta^2\right)\,\left(1+2\,\beta^2\right)\,x} ,\nonumber \\
T(x)&=&-\frac{\beta^2\,\left(\beta^2 + 2\,\beta^4 + x\,{\beta^2}'\right)}
{3\,\left(1+\beta^2\right)\,\left(1+2\,\beta^2\right)\,x^3\,q}.
\end{eqnarray}

It is the generalization of Eq.~(\ref{eq:pn_only}) to the general relativistic case.
The solution for the pressure, expressed in terms of the observed rotational velocity profile, $\beta(x)$, 
has the form
\begin{equation}
\bar{p}=\frac{\int{e^{\int^x{S(x')\,dx'}} T(x) dx + C}}{e^{\int^x{S(x')\,dx'}}}.
\label{eq:psol}
\end{equation}
where the value of the integration constant, $C$, is set by the appropriate boundary conditions.   

The density and the mass of the dark perfect fluid can be directly computed from 
Eqs.~(\ref{eq:nRG}-\ref{eq:pRG}), once the pressure is determined:
\begin{equation}
\bar{\rho}=\frac{\beta^2 + 2\,\beta^4 + x\,{\beta^2}'}{3\,\left(1+3\beta^2+\beta^4\right)\,x^2\,q}
- \frac{3 + 5\,\beta^2 - 2\,\beta^4 - 2\,x\,{\beta^2}'}{\left(1+3\beta^2+\beta^4\right)}\,p\,,\label{eq:rhosol}
\end{equation}
and
\begin{equation}
n=\frac{\beta^2}{\left(1+2\,\beta^2\right)\,q}\,x - 3\,\frac{x^3}{\left(1+2\,\beta^2\right)}\,p\,.
\label{eq:nsol}
\end{equation}

This is the exact treatment, the general relativistic one, to deal with the dark perfect fluid.
it provides the means to asses the accuracy of the Newtonian description. We also solved the cases
presented in this work within this exact treatment, and the results have an excellent agreement with 
the Newtonian ones, so that one can use such results with confidence.

\begin{figure}
\includegraphics[angle=0,width=.49\textwidth]{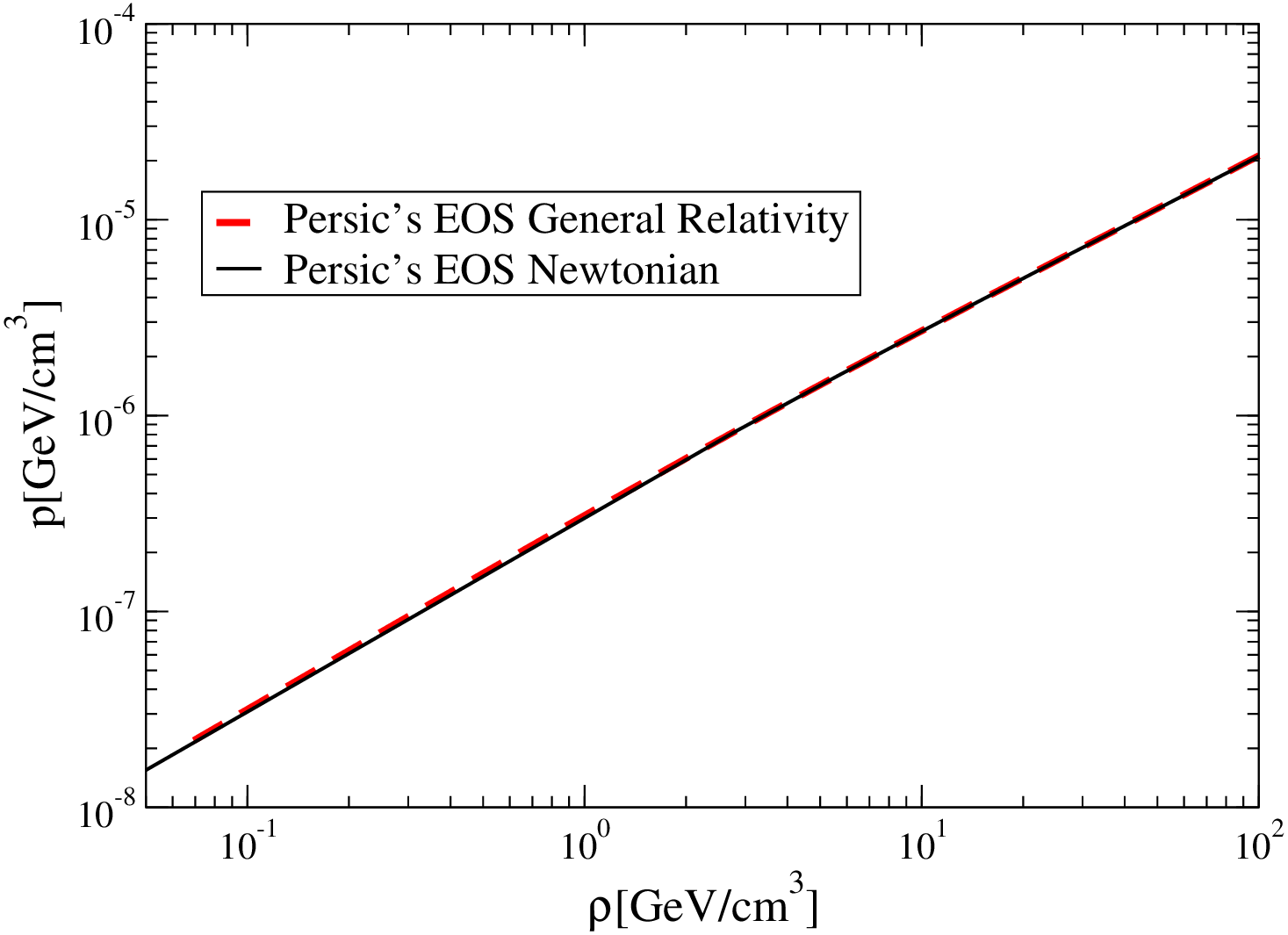}
\caption{Comparison of the resulting equation of state for the data points of the 
the galaxy U11748, the PSS's rotational velocity profile}\label{Fig12}
\end{figure}

In order to see the accuracy of the Newtonian treatment developed in all the article, we present a comparison
the General Relativistic results vs the Newtonian result for our well worked U11748 galaxy. See Fig.~(\ref{Fig12})

\label{lastpage}

\end{document}